\PassOptionsToPackage{pdfborder={0 0
0},colorlinks=true,linkcolor=blue,citecolor=blue}{hyperref}
\documentclass[compsoc]{IEEEtran}

\usepackage{graphicx}
\usepackage[tight,center]{subfigure}
\usepackage{booktabs}
\usepackage{amsmath}    
\usepackage{amssymb}
\usepackage{url}
\usepackage{array}
\usepackage{rotating}
\usepackage{multirow}
\usepackage{tabularx}
\usepackage{color}
\usepackage{colortbl}
\usepackage{todonotes}
\usepackage{soul}
\usepackage{util}
\usepackage{flushend}

\newif\ifdraft
\drafttrue

\newif\ifomegaOne
\omegaOnetrue

\def\Qthreshold{0.5}

\DeclareDocumentCommand\gfit{s}{\IfBooleanTF #1{\emph{Good Fits}}{\emph{Good Fit}}}
\DeclareDocumentCommand\ifit{s}{\IfBooleanTF #1{\emph{Inconclusive Fits}}{\emph{Inconclusive Fit}}}
\DeclareDocumentCommand\nfit{s}{\IfBooleanTF #1{\emph{Not Fits}}{\emph{Not Fit}}}

\def\odp{\ensuremath{\textit{os}}} 
\DeclareDocumentCommand\ap{sO{}}{\ensuremath{\textit{es}\IfBooleanTF #1{^*}{}_{#2}}} 
\def\DP{\ensuremath{\textit{OS}}}  

\DeclareDocumentCommand\AP{sO{}}{\ensuremath{\IfBooleanTF #1{\textit{ES}^*_{#2}}{\textit{ES}_{#2}} }}
\DeclareDocumentCommand\GAP{sO{}}{\ensuremath{\IfBooleanTF #1{\textit{GES}^*_{#2}}{\textit{GES}_{#2}} }}
\DeclareDocumentCommand\IAP{sO{}}{\ensuremath{\IfBooleanTF #1 {\textit{IES}^*_{#2}}{\textit{IES}_{#2}} }}
\DeclareDocumentCommand\NAP{sO{}}{\ensuremath{\IfBooleanTF #1 {\textit{NES}^*_{#2}}{\textit{NES}_{#2}} }}

\def\timespan{\ensuremath{\delta}}
\def\Timespan{\ensuremath{\Delta}}

\def\MSR{month}
\def\pvalue{\textit{p-value}}
\def\chisq{\ensuremath{\chi^2}}
\def\tick{\ensuremath{\bullet}}
\def\horizon{{\ensuremath{\tau}}}
\def\Horizon{{\ensuremath{\mathrm{T}}}}
\def\release{\ensuremath{r}}
\def\Release{\ensuremath{R}}
\newcommand\predict[1][\omega]{\ensuremath{\textit{Predict}_{#1}}}

\def\series#1{\ensuremath{{\footnotesize\textsf{TS}}(#1)}}
\newcommand\curve[2][vdm]{\ensuremath{\textit{#1}_{#2}}}
\newcommand\vdm[1][]{\ensuremath{\textit{vdm}_{#1}}}

\def\ds#1{{\ensuremath{\sf #1}}}

\def\ver#1{v#1}

\newcounter{step}
\renewcommand\thestep{Step \arabic{step}}

\newcommand\step{\refstepcounter{step}\thestep}

\newcounter{RQCounter}

\newcommand\researchquestion[2][border] {
\def\next{#1}
\refstepcounter{RQCounter}
\ifx\next\rqborder
    \fbox{
    \begin{description}
        \item [RQ\arabic{RQCounter}] \emph{#2}
    \end{description}}
\else
    \begin{description}
        \item [\bf RQ\arabic{RQCounter}] \emph{#2}
    \end{description}
\fi}

\newcounter{example}
\renewcommand\theexample{Example \arabic{example}}
\newenvironment{example}[1][]{%
\vspace{\baselineskip}
\refstepcounter{example}
\def\next{#1}
\noindent\textbf{\theexample\ifx\next\emptytext\else\;(#1)\fi} \hspace{2pt}
}{%
\hspace{\stretch{1}}\rule{1ex}{1ex} %
\vspace{\baselineskip}}

\newcounter{criteria}
\renewcommand\thecriteria{CR\arabic{criteria}}

\newif\ifusecolor
\ifusecolor
\definecolor{notfit}{rgb}{1,0,0}
\definecolor{notfit-text}{rgb}{0,0,0}
\definecolor{inclusive}{cmyk}{0,0,1,0}
\definecolor{fit}{rgb}{0,1,0}
\definecolor{aic-winner}{cmyk}{0,0,1,0}
\else
\definecolor{notfit}{gray}{0.25}
\definecolor{notfit-text}{gray}{1}
\definecolor{inclusive}{gray}{0.55}
\definecolor{fit}{gray}{0.85}
\definecolor{aic-winner}{gray}{0.70}
\fi

\def\gofN#1{\cellcolor{notfit}\textcolor{notfit-text}{$\times$}}
\def\gofI#1{\cellcolor{inclusive}\;}
\def\gofF#1{\cellcolor{fit}\checkmark}

\def\gofNT#1{\colorbox{notfit}{\textcolor{notfit-text}{\texttt{-}}}}
\def\gofIT#1{\colorbox{inclusive}{\texttt{?}}}
\def\gofFT#1{\colorbox{fit}{\texttt{X}}}

\newcommand\tableNVDGoF[2][]{
\begin{table}
    \centering
    \caption{#1}#2
    \ACCORCIA[-1.2]
    \setlength{\tabcolsep}{1pt}
    \extcaption[1\columnwidth]{
        The goodness of fit of a VDM is based on
        \emph{p-value} in the $\chi^2$ test.
        $\textit{p-value} < 0.05$: not fit (--),  $\textit{p-value} \geq
        0.95$: good fit (X), and inconclusive fit (?) otherwise.}

    \begin{tabularx}{1\columnwidth}{X*{17}{c}}
\toprule
 & \multicolumn{6}{c}{\textbf{Firefox}} & \multicolumn{6}{c}{\textbf{Chrome}} & \multicolumn{5}{c}{\textbf{IE}} \\
 \cmidrule(r){2-7} \cmidrule(r){8-13}  \cmidrule(r){14-18}
Model & 1.0 & 1.5 & 2.0 & 3.0 & 3.5 & 3.6 & 1.0 & 2.0 & 3.0 & 4.0 & 5.0 & 6.0 & 4.0 & 5.0 & 6.0 & 7.0 & 8.0 \\
\midrule
AML & \gofN{} & \gofN{} & \gofI{} & \gofI{} & \gofI{} & \gofI{} & \gofF{} & \gofI{} & \gofI{} & \gofI{} & \gofI{} & \gofI{} & \gofF{} & \gofI{} & \gofI{} & \gofN{} & \gofF{} \\
AT & \gofN{} & \gofN{} & \gofN{} & \gofN{} & \gofN{} & \gofN{} & \gofN{} & \gofN{} & \gofN{} & \gofN{} & \gofN{} & \gofN{} & \gofN{} & \gofN{} & \gofN{} & \gofI{} & \gofN{} \\
LN & \gofN{} & \gofN{} & \gofF{} & \gofN{} & \gofF{} & \gofI{} & \gofN{} & \gofN{} & \gofN{} & \gofI{} & \gofN{} & \gofN{} & \gofN{} & \gofN{} & \gofN{} & \gofI{} & \gofI{} \\
LP & \gofN{} & \gofN{} & \gofF{} & \gofI{} & \gofF{} & \gofF{} & \gofN{} & \gofN{} & \gofN{} & \gofN{} & \gofI{} & \gofI{} & \gofN{} & \gofF{} & \gofN{} & \gofF{} & \gofI{} \\
RE & \gofN{} & \gofN{} & \gofF{} & \gofI{} & \gofF{} & \gofF{} & \gofN{} & \gofN{} & \gofN{} & \gofN{} & \gofI{} & \gofI{} & \gofN{} & \gofF{} & \gofN{} & \gofI{} & \gofI{} \\
RQ & \gofN{} & \gofN{} & \gofN{} & \gofI{} & \gofI{} & \gofF{} & \gofN{} & \gofN{} & \gofI{} & \gofI{} & \gofI{} & \gofI{} & \gofN{} & \gofN{} & \gofN{} & \gofN{} & \gofF{} \\
\bottomrule
    \end{tabularx}
\end{table}
}

\renewcommand\tableNVDGoF[2][]{
\begin{table}
\caption{#1}#2
    \ACCORCIA[-1.2]
    \cmidrulekern 0.15em
    \extcaption[1\columnwidth]{
        The goodness of fit of a VDM is based on
        \pvalue\ in the $\chi^2$ test.
        $\pvalue < 0.05$: not fit ($\times$),  $\textit{p-value} \geq
        0.80$: good fit (\checkmark), and inconclusive fit (blank) otherwise. It is calculated over the entire lifetime.}
    \setlength{\tabcolsep}{1pt}
\resizebox{\columnwidth}{!}{\begin{tabular}{l*{30}{c}}
  \toprule
 & \multicolumn{8}{c}{\textbf{Firefox}} & \multicolumn{12}{c}{\textbf{Chrome}} & \multicolumn{5}{c}{\textbf{IE}} & \multicolumn{5}{c}{\textbf{Safari}} \\
  \cmidrule(lr){2-9} \cmidrule(lr){10-21}  \cmidrule(lr){22-26} \cmidrule(lr){27-31}
 & 1 & 1.5 & 2 & 3 & 3.5 & 3.6 & 4 & 5 & 1 & 2 & 3 & 4 & 5 & 6 & 7 & 8 & 9 & 10 & 11 & 12 & 4 & 5 & 6 & 7 & 8 & 1 & 2 & 3 & 4 & 5 \\
  \midrule
AML & \gofN{} & \gofN{} & \gofI{} & \gofN{} & \gofF{} & \gofI{} & \gofI{} & \gofI{} & \gofF{} & \gofI{} & \gofN{} & \gofI{} & \gofI{} & \gofI{} & \gofN{} & \gofI{} & \gofF{} & \gofI{} & \gofI{} & \gofN{} & \gofF{} & \gofI{} & \gofN{} & \gofI{} & \gofF{} & \gofN{} & \gofI{} & \gofI{} & \gofI{} & \gofF{} \\
  AT & \gofN{} & \gofN{} & \gofN{} & \gofN{} & \gofN{} & \gofN{} & \gofI{} & \gofI{} & \gofN{} & \gofN{} & \gofN{} & \gofN{} & \gofN{} & \gofN{} & \gofN{} & \gofN{} & \gofN{} & \gofN{} & \gofN{} & \gofI{} & \gofN{} & \gofN{} & \gofN{} & \gofN{} & \gofN{} & \gofN{} & \gofN{} & \gofN{} & \gofN{} & \gofI{} \\
  JW & \gofN{} & \gofN{} & \gofF{} & \gofF{} & \gofF{} & \gofF{} & \gofF{} & \gofI{} & \gofN{} & \gofN{} & \gofN{} & \gofN{} & \gofI{} & \gofI{} & \gofF{} & \gofF{} & \gofI{} & \gofI{} & \gofI{} & \gofI{} & \gofN{} & \gofF{} & \gofN{} & \gofF{} & \gofI{} & \gofN{} & \gofN{} & \gofN{} & \gofN{} & \gofN{} \\
  LN & \gofN{} & \gofN{} & \gofI{} & \gofN{} & \gofI{} & \gofI{} & \gofF{} & \gofI{} & \gofN{} & \gofN{} & \gofN{} & \gofN{} & \gofI{} & \gofI{} & \gofI{} & \gofI{} & \gofI{} & \gofI{} & \gofI{} & \gofN{} & \gofN{} & \gofN{} & \gofN{} & \gofF{} & \gofI{} & \gofN{} & \gofN{} & \gofN{} & \gofN{} & \gofI{} \\
  LP & \gofN{} & \gofN{} & \gofF{} & \gofF{} & \gofF{} & \gofF{} & \gofF{} & \gofI{} & \gofN{} & \gofN{} & \gofN{} & \gofN{} & \gofN{} & \gofI{} & \gofI{} & \gofI{} & \gofN{} & \gofI{} & \gofI{} & \gofI{} & \gofN{} & \gofF{} & \gofN{} & \gofN{} & \gofF{} & \gofN{} & \gofN{} & \gofN{} & \gofN{} & \gofI{} \\
  RE & \gofN{} & \gofN{} & \gofF{} & \gofF{} & \gofF{} & \gofF{} & \gofF{} & \gofI{} & \gofN{} & \gofN{} & \gofN{} & \gofN{} & \gofN{} & \gofI{} & \gofI{} & \gofI{} & \gofN{} & \gofI{} & \gofI{} & \gofI{} & \gofN{} & \gofF{} & \gofN{} & \gofN{} & \gofF{} & \gofN{} & \gofN{} & \gofN{} & \gofN{} & \gofI{} \\
  RQ & \gofN{} & \gofN{} & \gofN{} & \gofN{} & \gofN{} & \gofI{} & \gofF{} & \gofI{} & \gofN{} & \gofN{} & \gofN{} & \gofI{} & \gofI{} & \gofI{} & \gofI{} & \gofN{} & \gofN{} & \gofN{} & \gofN{} & \gofN{} & \gofN{} & \gofN{} & \gofN{} & \gofN{} & \gofF{} & \gofN{} & \gofN{} & \gofN{} & \gofN{} & \gofN{} \\
  YF & \gofN{} & \gofN{} & \gofF{} & \gofF{} & \gofF{} & \gofF{} & \gofF{} & \gofI{} & \gofF{} & \gofF{} & \gofF{} & \gofF{} & \gofF{} & \gofF{} & \gofF{} & \gofI{} & \gofN{} & \gofI{} & \gofI{} & \gofI{} & \gofF{} & \gofF{} & \gofF{} & \gofN{} & \gofF{} & \gofN{} & \gofN{} & \gofN{} & \gofN{} & \gofI{} \\
     \bottomrule
\end{tabular}}
\ACCORCIA
\end{table}
}

\begin{document}

\title{A Systematically Empirical Evaluation of Vulnerability Discovery Models: a Study on Browsers' Vulnerabilities }
\author{Viet~Hung~Nguyen and Fabio~Massacci}

\IEEEcompsoctitleabstractindextext{%
\begin{abstract}
A precise vulnerability discovery model (VDM) will provide a useful insight to
assess software security, and could be a good prediction instrument for both
software vendors and users to understand security trends and plan ahead
patching schedule accordingly. Thus far, several models have been proposed and
validated. Yet, no systematically independent validation by somebody other than
the author exists. Furthermore, there are a number of issues that might bias
previous studies in the field. In this work, we fill in the gap by introducing
an empirical methodology that systematically evaluates the performance of a VDM
in two aspects: quality and predictability. We further apply this methodology
to assess existing VDMs. The results show that some models should be rejected
outright, while some others might be adequate to capture the discovery process
of vulnerabilities. We also consider different usage scenarios of VDMs and find
that the simplest linear model is the most appropriate choice in terms of both
quality and predictability when browsers are young. Otherwise, logistics-based
models are better choices.
\end{abstract}

}

\maketitle


\section{Introduction}\label{sec:intro}
\PARstart{V}{ulnerability} discovery models (VDMs) operate on known
vulnerability data to estimate the total number of vulnerabilities that will be
reported after the release of a software. Successful models can be useful
instruments for both software vendors and users to understand security trends,
plan patching schedules, decide updates and forecast security investments in
general.

A VDM is a parametric mathematical function counting the number of cumulative
vulnerabilities of a software at an arbitrary time $t$. For example, if
$\Omega(t)$ is the cumulative number of vulnerabilities at time $t$, the
function of the linear model (LN) is  $\Omega(t) = At + B$ where $A, B$ are two
parameters of LN, which are calculated from the historical vulnerability data.

\begin{figure}
    \centering
    \includegraphics[width=\columnwidth,height=11\baselineskip]{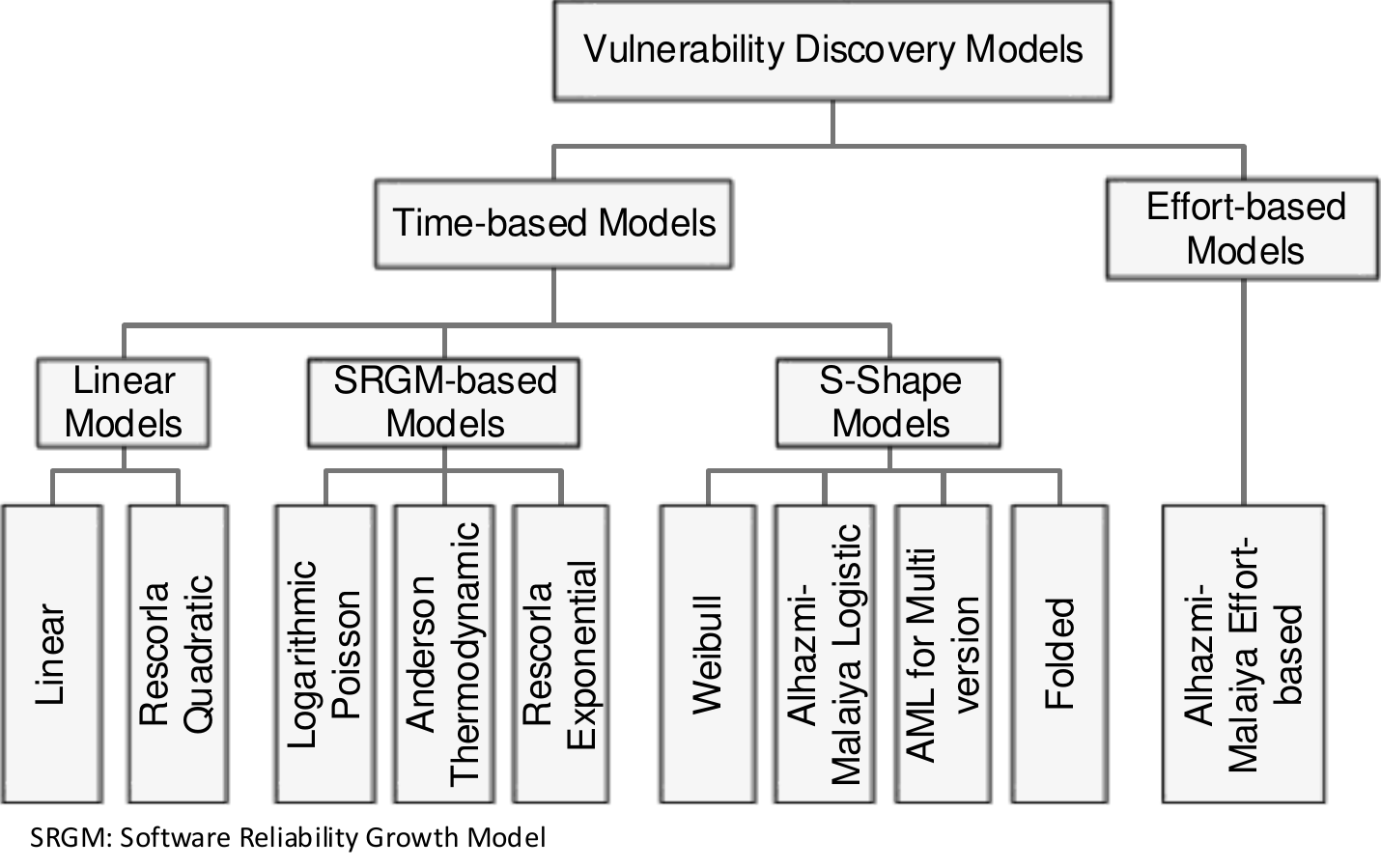}
    \caption{Taxonomy of Vulnerability Discovery Models.}
    \label{fig:VDM:taxonomy}
    \ACCORCIA
\end{figure}

\figref{fig:VDM:taxonomy} sketches a taxonomy of the major VDMs. It includes
Anderson's Thermodynamic (AT) model \cite{ANDE-02-OSS}, Rescorla's Quadratic
(RQ) and Rescorla's Exponential (RE) models \cite{RESC-05-SP}, Alhazmi \&
Malaiya's Logistic (AML) model \cite{ALHA-MALA-05-ISSRE}, AML for Multi-version
\cite{KIM-etal-07-HASE}, Weibull model (JW) \cite{JOH-etal-08-ISSRE}, and
Folded model (YF) \cite{YOUNIS-etal-11-SAM}. The \emph{goodness-of-fit} of
these models, \ie how well a model could fit the numbers of discovered
vulnerabilities, is normally evaluated in each paper on a specific
vulnerability data set, except AML which has been validated for different types
of application (\ie operating system \cite{ALHA-etal-05-DAS,ALHA-MALA-08-TR},
browsers \cite{WOO-ALHAZMI-MALAIYA-06-SEA}, web servers
\cite{ALHA-MALA-06-ISSRE,WOO-etal-11-CS}). Yet, no independent validation by
somebody other than the authors exists. Furthermore, a number of issues might
bias the results of previous studies.
\begin{itemize}
    \item Firstly, many studies did not clearly define what a vulnerability
        is. Indeed different definitions of vulnerability might lead to
        different counted numbers of vulnerabilities, and consequently,
        different conclusions.
    \item Secondly, all versions of a software were considered as a single
        ``entity". Even though there is a large amount of shared code, they
        are still different by a non-negligible amount of code.
    \item Thirdly, the goodness-of-fit of the models was often evaluated at
        a single time point (of writing their papers) and not used as a
        predictor \eg to forecast data for the next quarter for instance.
\end{itemize}
A detail discussion about these issues is available later in section
\secref{sec:questions}.

In this paper we want to address these shortcomings and derive a methodology
that can answer two basic questions concerning VDM: \emph{``Are VDMs adequate
to capture the discovery process of vulnerabilities?"}, and
\emph{``which VDM is the best?"}. 
%

\subsection{Contributions of This Paper}
The contributions of this work are detailed below:
\begin{itemize}
    \item We proposed an experimental methodology to assess the performance
        of a VDM based on its \emph{goodness-of-fit quality} and
        \emph{predictability}.
    \item We demonstrated the methodology by conducting an experiment
        analyzing eight VDMs, including AML, AT, JW, RQ, RE, LP, LN, and YF
        on 30 major releases of four popular browsers Internet Explorer
        (IE), Firefox (FF), Chrome and Safari.
    \item We presented an empirical evidence for the adequacy of the VDMs
        in terms of quality and predictability. The AT and RQ models are
        not adequate; whereas all other models may be adequate when
        software is young (12 months). However only s-shape models (AML,
        JW, YF) should be considered when software is middle age (36
        months) or older.
    \item We compared these VDMs (except AT and RQ) in different usage
        scenarios in terms of predictability and quality. The simplest
        model, LN, is more appropriate than other complex models when
        software is young and the prediction time span is not too long (12
        months or less). Otherwise, the AML model is superior. These
        results are summarized in \tabref{tbl:result}.
\end{itemize}

The rest of the paper is organized as follows. Section \secref{sec:terms}
presents terminology in our work. The research questions are presented in
section \secref{sec:questions} and the proposed methodology is described in
section \secref{sec:expdesign}. Next, we apply the methodology to analyze the
empirical performance of VDMs in all data sets in section \secref{sec:exp}.
Then in section \secref{sec:validity}, we discuss the threats to the validity
of our work. Finally, we review related work in section
\secref{sec:relatedwork}, and conclude in section \secref{sec:conclusion}.

\begin{table}
    \centering
    \caption{Performance summary of VDMs.}
    \label{tbl:result}
    \begin{tabularx}{\columnwidth}{cX}
        \toprule
        Model & Performance \\
        \midrule
        AT, RQ & should be rejected due to low quality. \\
        LN & is \emph{the best model} for first 12 months$^{(*)}$. \\
        AML & is \emph{the best model} for $13^{th}$ to $36^{th}$ month $^{(*)}$. \\
        RE, LP & may be adequate for first 12 months $^{(**)}$. \\
        JW, YF & may be adequate for $13^{th}$ to $36^{th}$ month$^{(*)}$. \\
        \bottomrule
    \end{tabularx}
    \ACCORCIA[-0.8]
    \extcaption[\columnwidth]{$^{(*)}$: in terms of quality and predictability for next 3/6/12 months. \\
    $^{(**)}$: in terms of quality and predictability for next 3 months.}
\end{table}

\section{Terminology}\label{sec:terms}
\begin{itemize}
    \item {\it A vulnerability } is ``an instance of a [human] mistake in
        the specification, development, or configuration of software such
        that its execution can [implicitly or explicitly] violate the
        security policy"\cite{KRSU-98-PHD}, later revised by
        \cite{OZMEN-07-QoP}. The definition covers all aspects of
        vulnerabilities discussed in
        \cite{ARBA-etal-00-IEEE,AVIZ-etal-04-TDSC,DOWD-etal-07,SCHNEIDER-91-NAP},
        see also \cite{OZMEN-07-QoP} for a discussion.
    \item {\it A data set} is a collection of vulnerability data extracted
        from one or more data sources.
    \item {\it A release} refers to a particular version of an application
        \eg Firefox \ver{1.0}.
    \item {\it A horizon} is a specific time interval sample. It is
        measured by the number of months since the released date, \eg from
        month $1$ to $12$ months after the release.
    \item {\it An observed vulnerability sample} (or observed sample, for
        short) is a time series of monthly cumulative vulnerabilities of a
        major release since the first month after release to a particular
        horizon.
    \item {\it An evaluated sample} is a tuple of an observed sample, a VDM
        model fitted to this sample (or another observed sample), and the
        goodness-of-fit of this model to this sample.
\end{itemize}

\section{Research Questions and Methodology Overview} \label{sec:questions}
In this work, we address the following two questions:

\researchquestion{How to evaluate the performance of a
VDM?}\label{rq:applicability}

\researchquestion{How to compare between two or more
VDMs?}\label{rq:comparison}

We propose a methodology to answer these questions. The proposed methodology
identifies data collection steps and mathematical analyses to empirically
assess different performance aspects of a VDM . The methodology is summarized
in \tabref{tbl:methodology}.

\begin{table*}
    \def\shade{\cellcolor[gray]{0.85}}
    \def\criteriashade{\cellcolor[gray]{0.95}}
    \caption{Methodology overview.}
    \label{tbl:methodology}
    \begin{tabularx}{\textwidth}{>{\sc}rX}
        \toprule
        \midrule
        \multicolumn{2}{l}{\shade\bf\step\label{step:data} Acquire the vulnerability
        data} \\
        desc. & Identify the vulnerability data sources, and the way to count vulnerabilities.
        If possible, different vulnerability sources should be used to select the most robust one.
        Observed samples then can be extracted from collected vulnerability
        data. \\
        input & Vulnerability data sources. \\
        output & Set of observed samples. \\
        criteria & \criteriashade{\bf\refstepcounter{criteria}\thecriteria\label{cr:dataset}}
                    \emph{Collection of observed samples}
            \begin{itemize}
                \item Vulnerabilities should be counted for individual
                    releases (possibly by different sources).
                \item Each observable sample should have at least $5$ data
                    points.
            \end{itemize} \\
        \midrule
        \multicolumn{2}{l}{\shade\bf\step\label{step:model-fit} Fit the VDM
        to observed
        samples} \\
        desc. & Estimate the parameters of the VDM formula to fit observed samples as much as possible.
        The \chisq\ goodness-of-fit test is employed to assess the
        goodness-of-fit of the fitted model based on criteria \ref{cr:gof}.\\
        input & Set of observed samples. \\
        output & Set of evaluated samples. \\
        criteria & \criteriashade{\bf\refstepcounter{criteria}\thecriteria\label{cr:gof}}
                {\it The classification of the evaluated samples based on the \pvalue\ of a \chisq\
                test.}
                \begin{itemize}
                    \item \textbf{Good Fit}: $\textit{p-value} \in [0.80, 1.0]$, a good
                        evidence to accept the model. We have more than $80\%$ chances of
                        generating the observed sample from the fitted model.
                    \item \textbf{Not Fit}: $\textit{p-value} \in [0, 0.05)$, a strong
                        evidence to reject the model. It means less than $5\%$ chances that the
                        fitted model would generate the observed sample.
                    \item \textbf{Inconclusive Fit}: $\textit{p-value} \in [0.05, 0.80)$,
                        there is not enough evidence to neither reject nor accept the
                        fitted model.
                \end{itemize}
        \\
        \midrule
        \multicolumn{2}{l}{\shade\bf\step\label{step:quality} Perform goodness-of-fit quality
        analysis} \\
        desc. & Analyze the goodness-of-fit quality of the fitted
        model by using the temporal quality metric which is the weighted ratio between fitted
        evaluated samples (both \gfit\ and \ifit) and total evaluated samples.  \\
        input & Set of evaluated samples. \\
        output & Temporal quality metric. \\
        criteria & \criteriashade{\bf\refstepcounter{criteria}\thecriteria\label{cr:quality}}
                {\it The rejection of a VDM.\newline}
        A VDM is rejected if it has a temporal quality lower than 0.5 even by counting
        \ifit* samples as positive (with weight 0.5). Different periods of software
        lifetime could be considered:
        \begin{itemize}
            \item $12$ months (young software)
            \item $36$ months (middle-age software)
            \item $72$ months (old software)
        \end{itemize}
        \\
        \midrule
        \multicolumn{2}{l}{\shade\bf\step\label{step:predictability} Perform predictability
        analysis} \\
        desc. & Analyze the predictability of the fitted model by
        using the predictability metric. Depending on different usage
        scenarios, we have different observation periods and time spans
        that the fitted model supposes to be able to predict. This is
        described in \ref{cr:timespan}.\\
        input & Set of evaluated samples. \\
        output & Predictability metric. \\
        criteria & \criteriashade
        {\refstepcounter{criteria}\bf\thecriteria\label{cr:timespan}}
            \emph{The observation period and prediction time spans based on some possible usage scenarios.}
            \newline\newline
            \resizebox{0.96\columnwidth}{!}{\begin{tabular}{lcc}
                 & Observation & Prediction \\
                Scenario & Period (months) & Time Span (months) \\
                \midrule
                Plan for short-term support & 6--24 & 3 \\
                Plan for long-term support & 6--24 & 12 \\
                Upgrade or keep & 6--12 & 6 \\
                Historic analysis & 24--36 & 12 \\
            \bottomrule
            \end{tabular}}
            \\
        \midrule
         \multicolumn{2}{l}{\shade\bf\step\label{step:comparison} Compare VDM} \\
         desc. & Compare the quality of the VDM with other VDMs by comparing their temporal quality and predictability metrics. \\
         input & Temporal quality and predictability measurements of models in comparison. \\
         output & Ranks of models. \\
         criteria & \criteriashade{\refstepcounter{criteria}\bf\thecriteria\label{cr:compare}}
         {\it The comparison between two VDM}
         \newline
          A VDM \vdm[1] is better than a VDM \vdm[2] if:
            \begin{itemize}
                \item either the predictability of \vdm[1] is significantly
                    greater than that of \vdm[2],
                \item or there is no significant difference between the
                    predictability of \vdm[1] and \vdm[2], but the temporal
                    quality of \vdm[1] is significantly greater than that
                    of \vdm[2].
            \end{itemize}
            The temporal quality and predictability should have their horizons and prediction time spans
            in accordance to criteria \ref{cr:quality} and \ref{cr:timespan}. Furthermore, a controlling procedure for multiple
            comparisons should be considered.
         \\
        \midrule
        \bottomrule
    \end{tabularx}
\end{table*}


In order to satisfactorily answer the questions above, we must address some
biases that potentially affected the validity of previous studies.

The \emph{vulnerability definition} bias may affect the vulnerability data
collection process. Indeed all previous studies reported their data sources,
but none clearly mentioned what a vulnerability is, and how to count it. A
vulnerability could be either an advisory reported by a software vendor such as
Mozilla Foundation Security Advisory -- MFSA, or a security bug causing
software to be exploited (reported in Mozilla Bugzilla), or an entry in
third-party vulnerability databases (\eg National Vulnerability Database --
NVD). Some entries may be classified differently by different entities: a
third-party database might report vulnerabilities, but the security bulletin of
vendors may not classify them as such. Consequently, the counted number of
vulnerabilities could be widely different depending on the different
definitions.

\begin{figure}
    \centering
    \includegraphics[width=0.7\columnwidth]{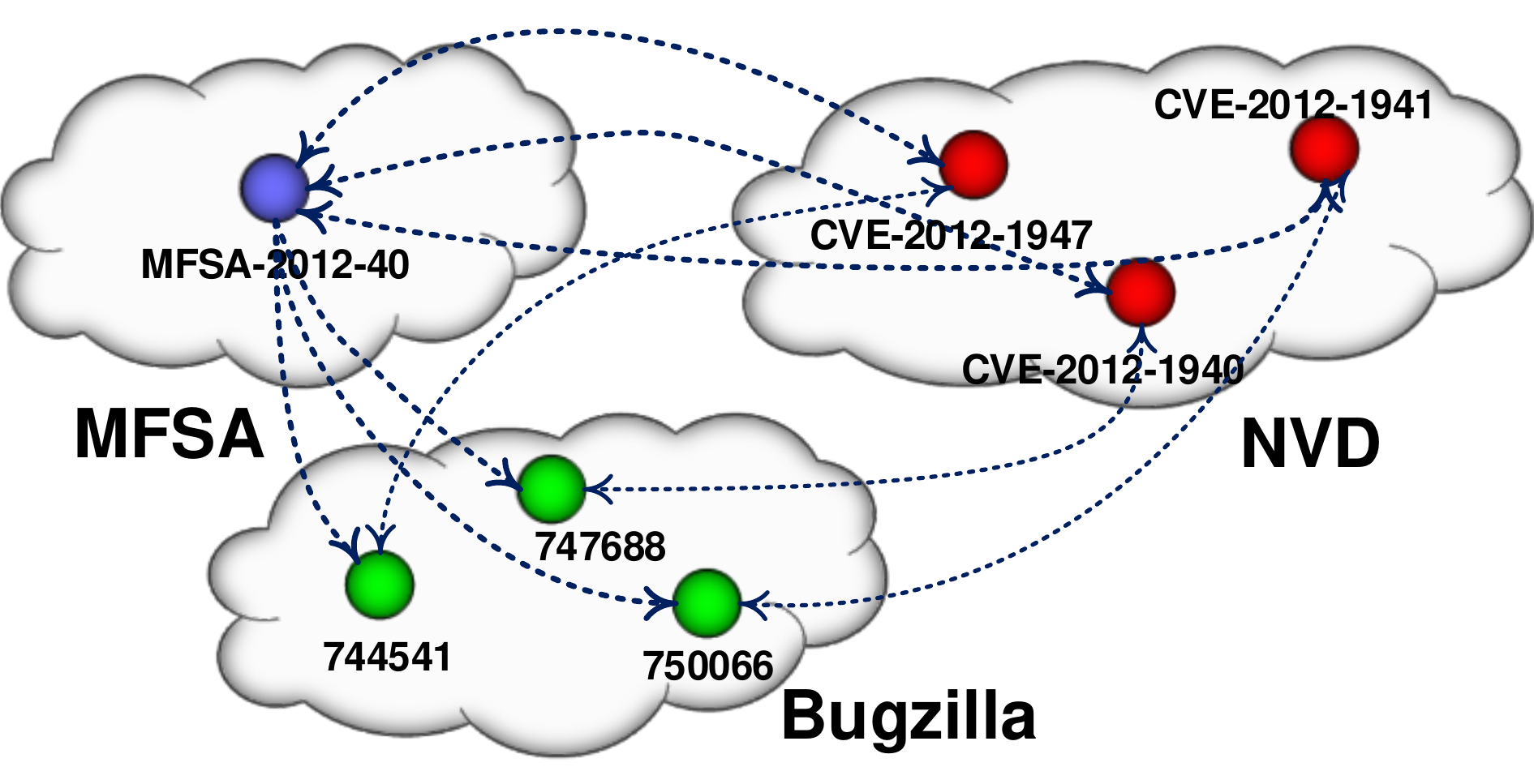}
    \extcaption[1\columnwidth]{A buffer-overflow flaw is reported in one entry of MFSA, three entries of
    Bugzilla, and three entries of NVD. There are cross-references among entries (arrow-headed
    dotted lines). So, how many vulnerabilities should we count?}
    \caption{The problem of counting vulnerabilities in Firefox.}
    \label{fig:issue:counting}
\end{figure}

\begin{example}
\figref{fig:issue:counting} exemplifies this issue. A security flaw concerning
the buffer overflow and use-after-free of Firefox \ver{13.0} is reported in
three databases with different number of entries: one MFSA entry
(\textsc{mfsa-2012-40}), three Bugzilla entries (\textsc{744541, 747688}, and
750066), and three NVD entries (\textsc{cve-2012-1947, cve-2012-1940}, and
\textsc{cve-2012-1941}). The cross references among these entries are
illustrated as directional connections. This figure raises a question ``how
many vulnerabilities should we count in this case?".
\end{example}

The \emph{multi-version software} bias affects the count of vulnerability
across releases. Some studies (\eg
\cite{WOO-ALHAZMI-MALAIYA-06-SEA,WOO-etal-11-CS}) considered all versions of
software as a single entity, and counted vulnerabilities for this entity. Our
previous study \cite{MASS-etal-11-ESSOS} has shown that each Firefox version
has its own code base, which may differ by $30\%$ or more from the immediately
preceding one. Therefore, as time goes by, we can no longer claim that we are
counting the vulnerabilities of the same application.

\begin{figure}[t]
    \centering
    \includegraphics[width=0.7\columnwidth]{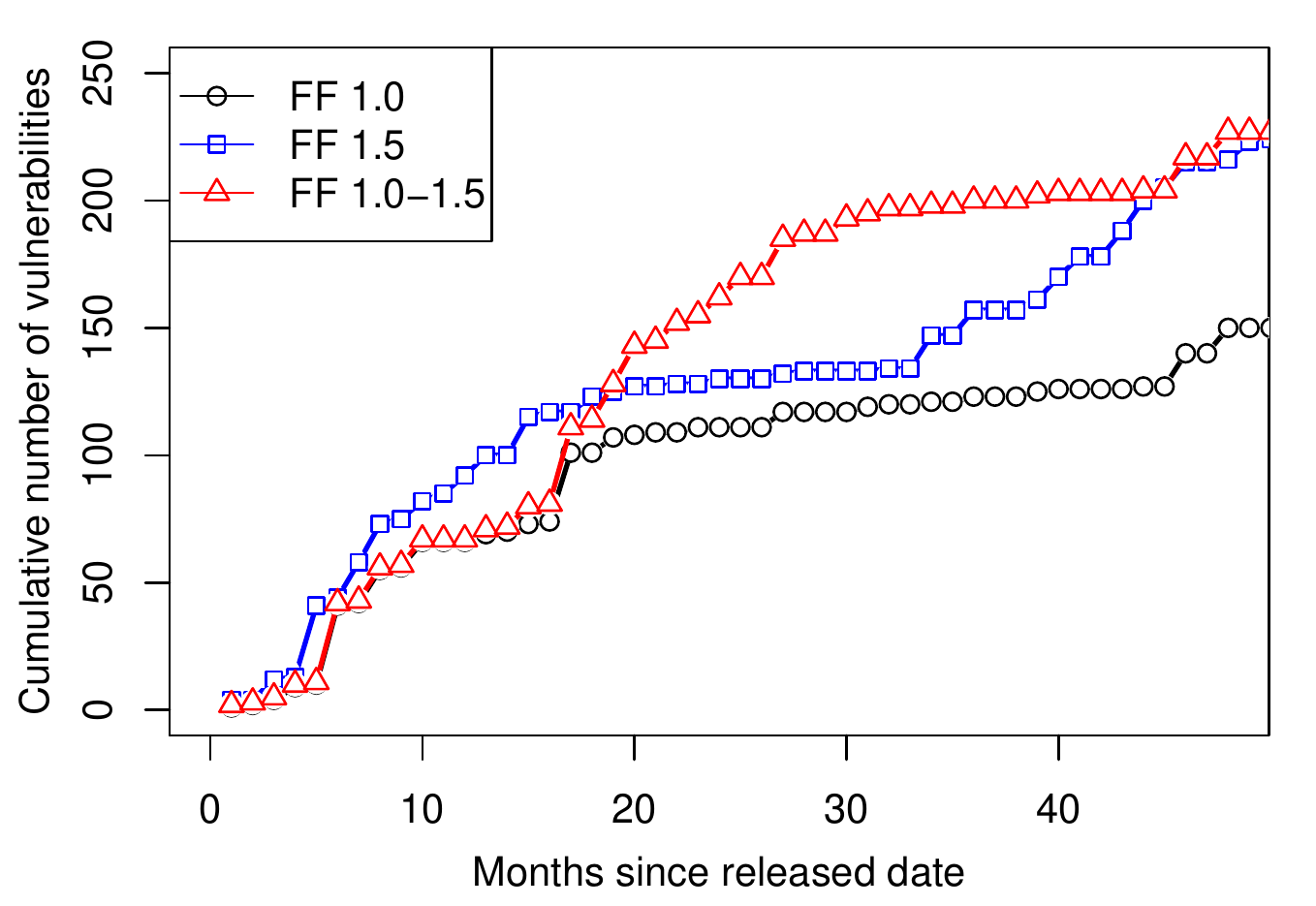}
    \extcaption[1\columnwidth]{This shows the cumulative vulnerabilities of Firefox \ver{1.0},
    \ver{1.5}, and \ver{1.0}-\ver{1.5} as a single entity. The ``global" entity exhibits
    a trend that is not present in the ``individual" versions.}
    \caption{Global vs individual vulnerability trends for Firefox.}
    \label{fig:firefox:firework}
    \ACCORCIA
\end{figure}

\begin{example}\label{ex:bias:multiversions}
\figref{fig:firefox:firework} visualizes this problem in a plot of the
cumulative vulnerabilities of Firefox \ver{1.0}, Firefox \ver{1.5}, and Firefox
\ver{1.0-1.5} as a single entity. Clearly, the function of the ``global"
version should be different from the functions of the individual versions.
\end{example}

The \emph{overfitting} bias, as name suggested, concerns the ability of a VDM
to explain history in hindsight. Previous studies took a snapshot of
vulnerability data, and fitted this entire snapshot to a VDM. This made a
brittle claim of fitness: the claim was only valid at the time vulnerabilities
were collected. It explained history but did not tell us anything about the
future. Meanwhile, we are interested in the ability of a VDM to be a good ``law
of nature" that is valid across releases and time and to some predict extent
the future.

\section{Methodology Details}\label{sec:expdesign}
This section discusses the details of our methodology to evaluate the
performance of a VDM.
\subsection{\ref{step:data}: Acquire the Vulnerability Data}\label{sec:step:data}
The acquisition of vulnerability data consists of two sub steps: \emph{Data set
collection}, and \emph{Data sample extraction}.

During \emph{Data set collection}, we identify the data sources to be used for
the study (as they may not equally fit to the task). We can classify them as
follows:
\begin{itemize}
    \item \emph{Third-party advisory} (\ds{TADV}): is a vulnerability
        database maintained by a third-party organization (not software
        vendors) \eg \ds{NVD}, Open Source Vulnerability Database
        (\ds{OSVDB}).
    \item \emph{Vendor advisory} (\ds{ADV}): is a vulnerability database
        maintained by a software vendor, \eg MFSA, Microsoft Security
        Bulletin. Vulnerability information in this database could be
        announced from third-party, but it is always validated before being
        announced as an advisory.
    \item \emph{Vendor bug tracker} (\ds{BUG}): is a bug-tracking database,
        usually maintained by vendors.
\end{itemize}

For our purposes, the following features of a vulnerability are interesting and
must be provided:
\begin{itemize}
    \item \emph{Identifier} (\ds{id}): is the identifier of a vulnerability
        within a data source.
    \item \emph{Disclosure date} (\ds{date}): refers to the date when a
        vulnerability is reported to the database\footnote{The actual
        discovery date might be significantly earlier than that.}.
    \item \emph{Vulnerable Releases} (\ds{\Release}):  is a list of
        releases affected by a vulnerability.
    \item \emph{References} (\ds{refs}): is a list of reference links to
        other data sources.
\end{itemize}
Not every feature is available from all data sources. To obtain missing
features, we can use \ds{id} and \ds{refs} to link across data sources and
extract the expected features from secondary data sources. 

\begin{example}
Vulnerabilities of Firefox are reported in three data sources:
NVD\footnote{Other third party data sources (\eg OSVDB, Bugtraq, IBM XForce)
also report Firefox's vulnerabilities, but most of them refer to NVD by the
CVE-ID. Therefore, we consider NVD as a representative of third-party data
sources.}, MFSA, and Mozilla Bugzilla.  Neither MFSA nor Bugzilla provides the
\emph{Vulnerable Releases} feature, but NVD does. Each MFSA entry has one or
more links to NVD and Bugzilla. Therefore, we could to combine MFSA and NVD,
Bugzilla and NVD to obtain the missing data. 
\end{example}


We address the \emph{vulnerability definition} bias by taking into account
different definitions of vulnerability. Particularly, we collected different
vulnerability data sets with respect to these definitions. We also address the
\emph{multi-version} issue by collecting vulnerability data for individual
releases. \tabref{tbl:datasets:definition} shows different data sets that we
have considered in our study. They are combinations of three types of data
sources : third-party (\ie NVD as a representative), vendor advisory, and
vendor bug tracker. The English descriptions of these data sets for a
\emph{release} $r$ are as follows:
\begin{itemize}
    \item \ds{NVD(\emph{r})}: a set of NVD entries which claim $r$ is
        vulnerable.
    \item  \ds{NVD.Bug(\emph{r})}: a set of NVD entries which are confirmed
        by at least a vendor bug report, and claim $r$ is vulnerable.
    \item \ds{NVD.Advice(\emph{r})}: a set of NVD entries which are
        confirmed by at least a vendor advisory, and claim $r$ is
        vulnerable. Notice that the advisory report might \emph{not}
        mention $r$, but later releases.
    \item \ds{NVD.Nbug(\emph{r})}: a set of vendor bug reports confirmed by
        NVD, and $r$ is claimed vulnerable by NVD.
    \item \ds{Advice.NBug(\emph{r})}: a set of bug reports mentioned in an
        advisory report of a vendor. The advisory report also refers to at
        least an \ds{NVD} entry that claims $r$ is vulnerable.
\end{itemize}

\begin{table}
    \centering
    \caption{Formal definition of data sets.}
    \label{tbl:datasets:definition}
    \resizebox{\columnwidth}{!}{
    \begin{tabular}{l>{$}l<{$}}
        \toprule
        Data set & \text{Definition} \\
        \midrule
        NVD(\release) & \Set{\ds{nvd \in \ds{NVD}| \release \in \Release_{nvd}}} \\
        NVD.Bug(\release) & \Set{\ds{nvd \in \ds{NVD}|
                    \exists b \in BUG: \release \in \Release_{nvd} \wedge id_\ds{b} \in refs_\ds{nvd}}} \\
        NVD.Advice(\release) & \Set{\ds{nvd \in \ds{NVD}|
                    \exists a \in ADV: \release \in \Release_{nvd} \wedge id_\ds{a} \in refs_\ds{nvd}}} \\
        NVD.NBug(\release) & \Set{\ds{b \in BUG|
                    \exists nvd \in NVD: \release \in \Release_{nvd} \wedge id_\ds{b} \in refs_\ds{nvd}}} \\
        Advice.NBug(\release) & \left\{\ds{b \in BUG|
                    \exists a \in ADV,\exists nvd \in NVD: \release \in \Release_{nvd}}\right. \\
                  & \left.\ds{\wedge id_{b} \in refs_{a} \wedge id_{nvd} \in refs_{a} \wedge
                  \operatorname{cluster}_a(id_b, id_{nvd})}\right\}\\
        \bottomrule
    \end{tabular}}
    \ACCORCIA[-0.8]
    \extcaption[1\columnwidth]{\emph{Note}: $\ds{\Release_{nvd},
    refs_\ds{nvd}}$ denote the vulnerable releases and references of an entry
    \ds{nvd}, respectively. $\ds{id_\ds{a}, id_\ds{b}, id_{nvd}}$ denote the identifier
    of \ds{a}, \ds{b}, and \ds{nvd}.
    $\operatorname{cluster}_\ds{a}(\ds{id_b,id_{nvd}})$ is a predicate
    checking whether $\ds{id_b}$ and $\ds{id_{nvd}}$ are located next together in the advisory $\ds{a}$.}
    \ACCORCIA[-1.5]
\end{table}

For \emph{Data sample extraction}, we extract observed samples from collected
data sets. An \emph{observed sample} is a time series of (monthly) cumulative
vulnerabilities of a release. It starts from the first month since release to
the end month, called \emph{horizon}. A month is an appropriate granularity for
sampling because week and day are too short intervals and are subject to random
fluctuation. Additionally, this granularity was used in the literature.

Let \Release\ be the set of analyzed releases and $DS$ be the set of data sets,
an observed sample (denoted as $\odp$) is a time series defined as follows:
\begin{equation}
    \odp = \series{\release, ds, \horizon}
\end{equation}
where:
\begin{itemize}
    \item $\release \in \Release$ is a release in the evaluation;
    \item $ds \in DS$ is the data set where samples are extracted;
    \item $\horizon \in \Horizon_\release = \left[\horizon^\release_{min},
        \horizon^\release_{max}\right]$ is the horizon of the observed
        sample, in which $\Horizon_\release$ is the \emph{horizon range of
        release \release}.
\end{itemize}
In the horizon range of release \release, the minimum value of horizon
$\horizon^\release_{min}$ of \release\ depends on the starting time of the
first observed sample of \release. Here we choose $\horizon^\release_{min}=6$
for all releases so that all observed samples have enough data points for
fitting all VDMs. The maximum value of horizon $\horizon^\release_{max}$
depends on how long the data collection period is for each release.

\begin{example}
IE \ver{4.0} was released in September, $1997$\footnote{Wikipedia,
\url{http://en.wikipedia.org/wiki/Internet_Explorer}, visited on 24 June
2012.}. The first \MSR\ was on $31$ October, $1997$. The first observed sample
of IE \ver{4.0} is a time series of $6$ numbers of cumulative vulnerabilities
for the $1^{st},2^{nd},\ldots,6^{th}$ months. Since the date of data collection
is on 01 July 2012, IE \ver{4.0} have been released for 182 months, and
therefore has 177 observed samples. Hence the maximum value of horizon
($\horizon^\ds{IE\ver{4.0}}_{max}$) is $182$.
\end{example}

\subsection{\ref{step:model-fit}: Fit a VDM to Observed
Samples}\label{sec:step:modelfit}

We estimate the parameters of the VDM formula by a regression method so that
the VDM curve fits an observed sample as much as possible. We denote the fitted
curve (or fitted model) as:
\begin{equation}
\curve{\series{\release, ds,\horizon}} \label{eq:model}
\end{equation}
where $vdm$ is the VDM being fitted; $\odp=\series{\release, ds, \horizon}$ is
an observed sample from which the \vdm's parameters are estimated.
\eqref{eq:model} could be shortly written as \curve{\odp}.

\begin{figure*}
    \centering
    \includegraphics[width=0.9\textwidth]{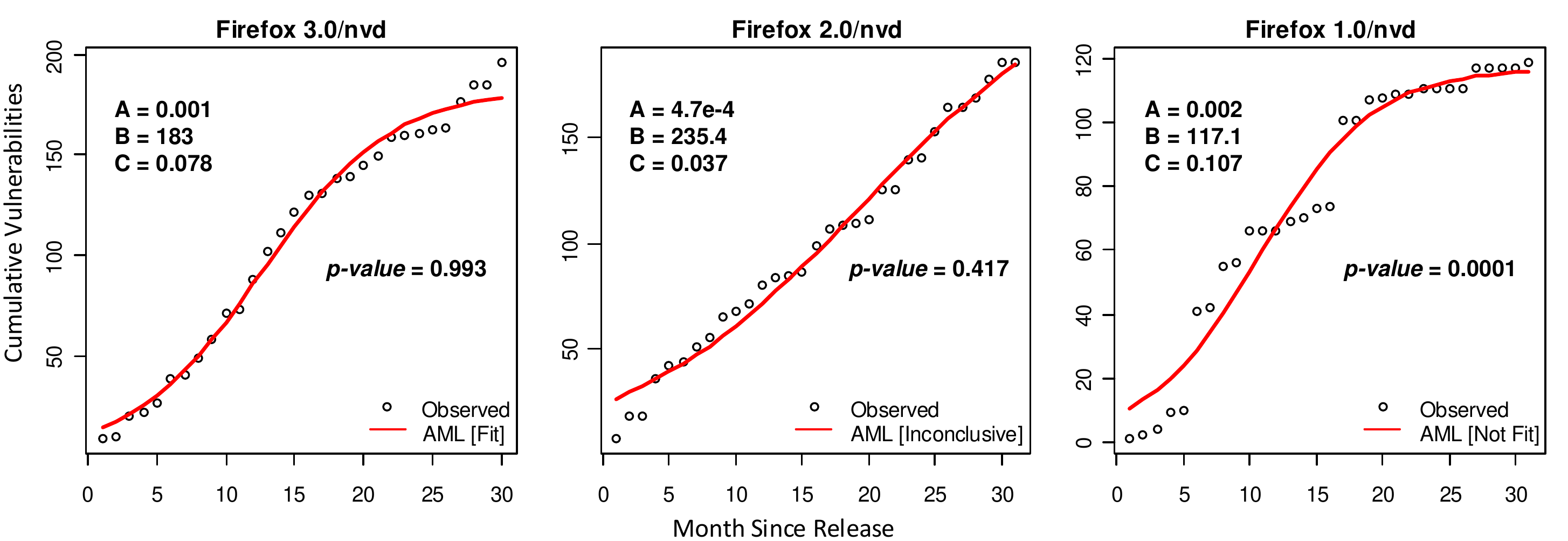}
    \extcaption[0.8\textwidth]{A,B,C are three parameters in the formula of the AML model: $\Omega(t) = \frac{B}{BCe^{-ABt} + 1}$ (see also \tabref{tbl:vdm})}
    \caption{Fitting the AML model to the \ds{NVD} data sets for Firefox \ver{3.0}, \ver{2.0}, and
    \ver{1.0}.}
    \label{fig:vdm:fit}
\end{figure*}

\begin{example}
Fitting the AML model to the NVD data set of Firefox \ver{3.0} at the $30^{th}$
\MSR, \ie the observed sample $\odp=\series{\ds{FF3.0},\ds{NVD}, 30}$,
generates the curve:
\[
    \curve[AML]{\series{\ds{FF3.0}, \ds{NVD}, 30}} = \frac{183}{183\cdot0.078\cdot e^{-0.001 \cdot 183 \cdot t} + 1}
\]
\figref{fig:vdm:fit} illustrates the plots of three curves
\curve[AML]{\series{\release, \ds{NVD}, 30}}, where $r$ is \ds{FF3.0, FF2.0},
and \ds{FF1.0}. The X-axis is the number of months since release, and the
Y-axis is the cumulative number of vulnerabilities. Circles represent observed
vulnerabilities. The solid line indicates the fitted AML curve.
\end{example}

In \figref{fig:vdm:fit}, the distances of the circles to the curve are used to
estimate the goodness-of-fit of the model. The goodness-of-fit is measured by
the Pearson's Chi-Square ($\chi^2$) test, which is a common test in the
literature. In this test, we measure the \chisq\ statistic value of the curve
by using the following formula:
\begin{equation}
    \chisq = \sum_{t=1}^{\horizon}\frac{(O_t - E_t)^2}{E_t} \label{eq:chisq}
\end{equation}
where $O_t$ is the observed cumulative number of vulnerabilities at time $t$
(\ie $t^\textnormal{th}$ value of the observed sample); $E_t$ denotes the
expected cumulative number of vulnerabilities which is the value of the curve
at time $t$. The \chisq\ value is proportional to the differences between the
observed values and the expected values. Hence, the larger \chisq, the smaller
goodness-of-fit. If the \chisq\ value is large enough, we can safely reject the
model. In other words, the model statistically does not fit the observed data
set. The \chisq\ test requires all expected values should be at least $5$ to
ensure the validity of the test \cite[Chap. 1]{NIST-StatBook-12}. If there is
any expected value is less than 5, we need to combine some first months to
increase the expected value \ie increase the starting value of $t$ in
\eqref{eq:chisq} until $E_t \ge 5$.


The conclusion about whether a VDM curve statistically fits an observed sample
relies on the \pvalue\ of the test, which is derived from \chisq\ value and the
degrees of freedom (\ie the number of months minus one). Semantically, the
\pvalue\ is the probability that we falsely reject the \emph{null hypothesis}
when it is true (\ie error Type I: false positive). The null hypothesis here
is: \emph{``there is no statistical difference between observed and expected
values."} which means that the model fits the observed sample. Therefore, if
the \pvalue\ is less than the significance level $\alpha$ of $0.05$, we can
reject a VDM because there is less than $5\%$ chances that this fitted model
would generate the observed sample.

In contrast, to accept a VDM, we exploit the power of the \chisq\ test which is
the probability of rejecting the null hypothesis when it is false. Normally,
`an $80\%$ power is considered desirable' \cite[Chap. 8]{MCKILLUP-BOOK}. Hence
we accept a VDM if the \pvalue\ is greater than or equal to $0.80$. We have
more than $80\%$ chances of generating the observed sample from the fitted
curve. In all other cases, we should neither accept nor reject the model
(inconclusive fit).

The criteria \ref{cr:gof} in \tabref{tbl:methodology} summarizes the way by
which we assess the goodness-of-fit of a fitted model based on the \pvalue\ of
the \chisq\ test.

In the sequel, we use the term \emph{evaluated sample} to denote the triplet
composed by an observed sample, a fitted model, and the \pvalue\ of the \chisq\
test.

\begin{example}
In \figref{fig:vdm:fit}, the first plot shows the AML model with a \emph{Good
Fit} ($\pvalue = 0.993 > 0.80$), the second plot exhibits the AML model with an
\emph{Inconclusive Fit} ($0.05 < \pvalue = 0.417 < 0.80$), and the last one
denotes the AML model with a \emph{Not Fit} ($\pvalue=0.0001 < 0.05$).
\end{example}


There are also other statistic tests for goodness-of-fit, for instance the
Kolmogorov-Smirnov (K-S) test, and the Anderson-Darling (A-D) test. The K-S
test is an exact test; it, however, only applies to continuous distributions.
An important assumption is that the parameters of the distribution cannot be
estimated from the data. Hence, we cannot apply it to perform the
goodness-of-fit test for a VDM. The A-D test is a modification of the K-S test
that works for some distributions \cite[Chap. 1]{NIST-StatBook-12} (\ie normal,
log-normal, exponential, Weibull, extreme value type I, and logistic
distribution), but some VDMs violate this assumption.

\subsection{\ref{step:quality}: Perform Goodness-of-Fit Quality Analysis}\label{sec:step:quality}
To address the \emph{overfitting} bias, we introduce the \emph{goodness-of-fit
quality} (or \emph{quality}, for short) that measures the overall number of
\emph{Good Fit}s and \emph{Inconclusive Fit}s among different samples. In
contrast, previous studies considered only one observed sample which is the one
with the largest horizon in their experiment.

Let $\DP = \Set{\series{\release, ds, \horizon}|\release \in R \wedge ds \in DS
\wedge \horizon \in \Horizon_\release}$ be the set of observed samples, the
\emph{overall quality} of a model \vdm\ is defined as the weighted ratio of the
number of \emph{Good Fit} and \emph{Inconclusive Fit} evaluated samples over
the total ones, as shown bellow:
\begin{equation}
    Q_\omega = \frac{|\GAP| + \omega \cdot |\IAP|}{|\AP|} \label{eq:quality:global}
\end{equation}
where:
\begin{itemize}
    \item $\AP = \Set{\Seq{\odp, \curve{\odp}, p}|\odp \in \DP}$ is the set
        of evaluated samples generated by fitting \vdm\ to observed
        samples;
    \item $\GAP = \Set{\Seq{\odp, \curve{\odp}, p} \in \AP|p \ge 0.80}$ is
        the set of \emph{Good Fit} evaluated samples;
    \item $\IAP = \Set{\Seq{\odp, \curve{\odp}, p} \in \AP| 0.05 \le p <
        0.80}$ is the set of \emph{Inconclusive Fit} evaluated samples;
    \item  $\omega \in [0..1]$ is the \emph{inconclusiveness contribution}
        factor denoting that an \emph{Inconclusive Fit} is $\omega$ times
        less important than a \emph{Good Fit}.
\end{itemize}

\begin{example}
If we fit the AML model to $3,895$ observed samples of the four browsers IE,
Firefox, Chrome, and Safari. For $1,526$ times AML is a \gfit, and for $1,463$
times AML is an \ifit. The overall quality of AML is:
\begin{align*}
    Q_{\omega=0} &= \frac{1,526}{3,895} = 0.39 \\
    Q_{\omega=1} &= \frac{1,526 + 1,463}{3,895} = 0.77 \\
    Q_{\omega=0.5} &= \frac{1,526 + 0.5 \cdot 1,463}{3,895} = 0.58
\end{align*}
To calculate the \chisq\ test we refit the model each and every time. So this
means that we have $1,526$ different parameters A, B and C for each good fit
curve (see \figref{fig:vdm:fit}).
\end{example}

The overall quality metric ranges between 0 and 1. The quality of 0 indicates a
completely inappropriate model, whereas the quality of 1 indicates a perfect
one. This metric is a very optimistic measure as we are essentially
``refitting" the model as more data become available. Hence, it is the upper
bound value of the VDM quality.

The factor $\omega$ denotes the contribution of an inconclusive fit to the
overall quality. A skeptical analyst would expect $\omega=0$, which means only
\emph{Good Fits} are meaningful. Meanwhile an optimistic analyst would set
$\omega=1$, which mean an \emph{Inconclusive Fit} is as good as a \emph{Good
Fit}. The optimistic choice $\omega=1$ is usually adopted by the proposers of
each model in previous studies in the field while assessing the VDM quality.
The effect of the $\omega$ factor on the overall quality metrics is illustrated
in \figref{fig:quality:omega} showing the variation of the overall quality of
two models AML and AT with respect to $\omega$. We do not know whether an
\ifit\ is good or not because the observed samples do not provide enough
evidence. Hence, the choice of $\omega = 0.5$ may be considered a good balance.
During our analysis we use $\omega = 0.5$; any exception will be explicitly
noted.

\begin{figure}
    \centering
    \includegraphics[width=0.7\columnwidth]{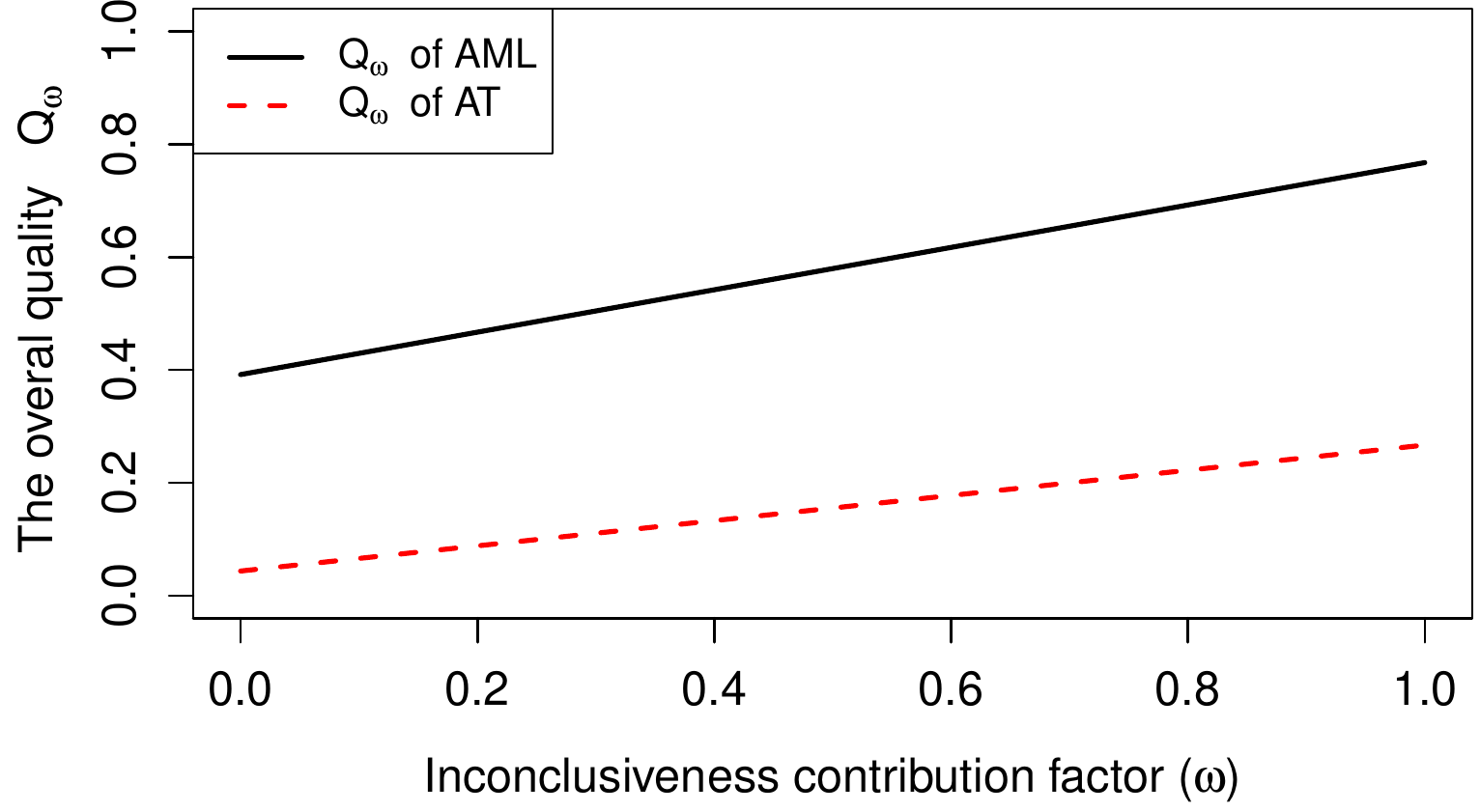}
    \caption{The variation of the overall quality $Q_\omega$ with respect to the $\omega$ factor.}
    \label{fig:quality:omega}
\end{figure}


The overall quality metric is sensitive to brittle performance in time. A VDM
could produce a lot of \gfit* evaluated samples for the first $6$ months, but
almost \nfit* at other horizons. Unfortunately, the metric did not address this
phenomenon.


To avoid this unwanted effect, we introduce the \emph{temporal quality} metric
which represents the evolution of the overall quality over time. The temporal
quality $Q_\omega(\horizon)$ is the weighted ratio of the \emph{Good Fit} and
\emph{Inconclusive Fit} evaluated samples over total ones at the particular
horizon \horizon. The temporal quality is formulated in the following equation:
\begin{equation}
    Q_\omega(\horizon) = \frac{|\GAP(\horizon)| + \omega \cdot |\IAP(\horizon)|}{|\AP(\horizon)|}
    \label{eq:quality:horizon}
\end{equation}
where:
\begin{itemize}
    \item $\horizon \in \Horizon$ is the horizon that we observe samples,
        in which $\Horizon \subseteq \bigcup_{\release \in
        \Release}\Horizon_\release$ is the subset of the union of the
        horizon ranges of all releases \release\ in evaluation;
    \item $\AP(\horizon) = \Set{\Seq{\odp,\curve{\odp},p}|\odp \in
        \DP(\horizon)}$ is the set of evaluated samples at the horizon
        $\horizon$; where \DP(\horizon) is the set of observed samples at
        the horizon \horizon\ of all releases; 
    \item $\GAP(\horizon) \subseteq \AP(\horizon)$ is the set of \emph{Good
        Fit} evaluated samples at the horizon \horizon;
    \item $\IAP(\horizon) \subseteq \AP(\horizon)$ is the set of
        \emph{Inconclusive Fit} evaluated samples at the horizon \horizon;
    \item $\omega$ is the same as for the overall quality $Q_\omega$.
\end{itemize}

To study the trend of the temporal quality $Q_\omega(\horizon)$, we employ the
\emph{moving average} technique which is commonly used in time series analysis
to smooth out short-term fluctuations and highlight longer-term trends.
Intuitively each point in the moving average is the average of some adjacent
points in the original series.  The moving average of the temporal quality is
defined as follows:
\begin{equation}
    \textit{MA}_k^{Q_\omega}(\horizon) = \frac{1}{k}\sum_{i=1}^kQ_\omega(\horizon - i + 1)\label{eq:ma:q}
\end{equation}
where $k$ is the \emph{window size}. The choice of $k$ changes the
spike-smoothening effect: higher $k$, smoother spikes. Additionally, $k$ should
be an odd number so that variations in the mean are aligned with variations in
the data rather than being shifted in time.

\begin{figure}[t]
    \centering
    \includegraphics[width=1\columnwidth]{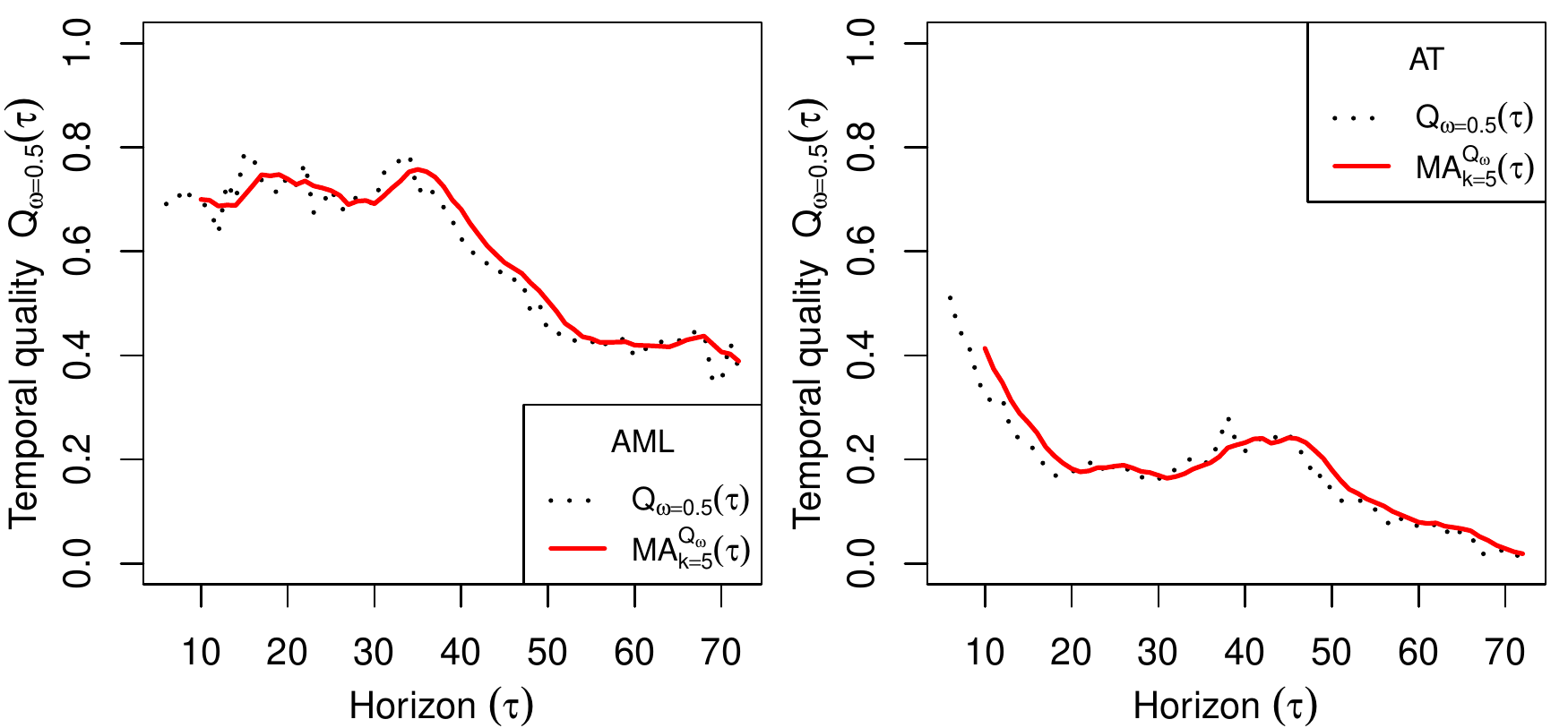}
    \extcaption[1\columnwidth]{Dotted lines are the temporal quality with $\omega=0.5$, solid lines
    are the moving average of the temporal quality with the window size $k=5$.}
    \caption{An example about the moving average of the temporal quality of AML and AT models. }
    \label{fig:quality:ma:example}
\end{figure}

\begin{example}
\figref{fig:quality:ma:example} depicts the moving average for the temporal
quality of AML and AT models. In this example, we choose a window size $k=5$
because the minimum horizon is six ($\horizon^\release_{min} = 6$), so $k$
should be less than this horizon ($k < \horizon^\release_{min}$); and $k=3$ is
too small to smooth out the spikes. \ACCORCIA
\end{example}

\subsection{\ref{step:predictability}: Perform Predictability Analysis}\label{sec:step:predictability}
The predictability of a VDM measures the capability of predicting future trends
of vulnerabilities. This essentially makes a VDM applicable in practice. The
calculation of the predictability of a VDM has two phases, the \emph{learning
phase} and the \emph{prediction phase}. In the learning phase, we fit a VDM to
an observed sample at a certain horizon. In the prediction phase, we evaluate
the qualities of the fitted model on observed samples in future horizons.


We extend \eqref{eq:quality:horizon} to calculate the prediction quality. Let
\curve{\series{\release,ds,\horizon}} be a fitted model at horizon \horizon.
The prediction quality of this model in the next \timespan\ months is
calculated as follows:
\begin{equation}
     Q^*_\omega(\horizon,\timespan) = \frac{|\GAP*(\horizon,\timespan)| + \omega \cdot
     |\IAP*(\horizon,\timespan)|}{|\AP*(\horizon,\timespan)|} \label{eq:quality:predict}
\end{equation}
where:
\begin{itemize}
    \item $\AP*(\horizon,\timespan) =
        \Set{\Seq{\series{\release,ds,\horizon+\timespan},
        \curve{\series{\release,ds,\horizon}}, p}}$ is the set of evaluated
        samples at the horizon $\horizon+\timespan$ in which we evaluate
        the quality of the model fitted at horizon \horizon\
        (\curve{\series{\release,ds,\horizon}}) on observed samples at the
        future horizon $\horizon + \timespan$. We refer to
        $\AP*(\horizon,\timespan)$ as set of evaluated samples of
        prediction;
    \item $\GAP*(\horizon,\timespan) \subseteq \AP*(\horizon,\timespan)$ is
        the set of \emph{Good Fit} evaluated samples of prediction at the
        horizon $\horizon + \timespan$;
    \item $\IAP*(\horizon,\timespan) \subseteq \AP*(\horizon,\timespan)$ is
        the set of \emph{Inconclusive Fit} evaluated samples of prediction
        at the horizon $\horizon + \timespan$.
    \item $\omega$ is the same as for the overall quality $Q_\omega$.
\end{itemize}



\begin{example}
\figref{fig:predict:quality} illustrates the prediction qualities of two models
AML and AT starting from the horizon of $12^{th}$ month ($\horizon=12$, left)
and $24^{th}$ month ($\horizon=24$, right), and predicting the value for next
$12$ months ($\timespan=0 \ldots 12$). White circles are prediction qualities
of AML, and red (gray) circles are those of AT.
\end{example}


As we can see from \figref{fig:predict:quality} the ability of a  model to
predict data decreases with time. It is therefore useful to identify some
interesting prediction time spans (such as next $6$ months) that can be used
for pairwise comparison between VDMs. To this extent, we identify some
different scenarios by which we specify the duration of data observation and
the prediction time span. Other scenarios may be identified depending on the
application or the readers' interest:
\begin{itemize}
    \item \emph{Plan for short-term support}: the data observation period
        may vary from $6$ months to the whole lifetime. We are looking for
        the ability to predict the trend in next quarter (\ie $3$ months)
        to plan the short-term support activities \eg allocating resources
        for fixing vulnerabilities.
    \item \emph{Plan for long-term support}: we would like to predict a
        realistic expectation for bug reports in the next 1 year to plan
        the long-term activities.
    \item \emph{Upgrade or keep}: the data observation period is short (at
        from 6 to 12 months). We are looking on what is going to happen in
        next 6 months. For example to decide whether to keep the current
        system or to go over the hassle of updating it.
    \item \emph{Historic analysis}: the data observation period is long (2
        to 3 years), we are considering what happens for extra support in
        the next 1 year.
\end{itemize}

\begin{figure}
    \centering
    \includegraphics[width=0.98\columnwidth]{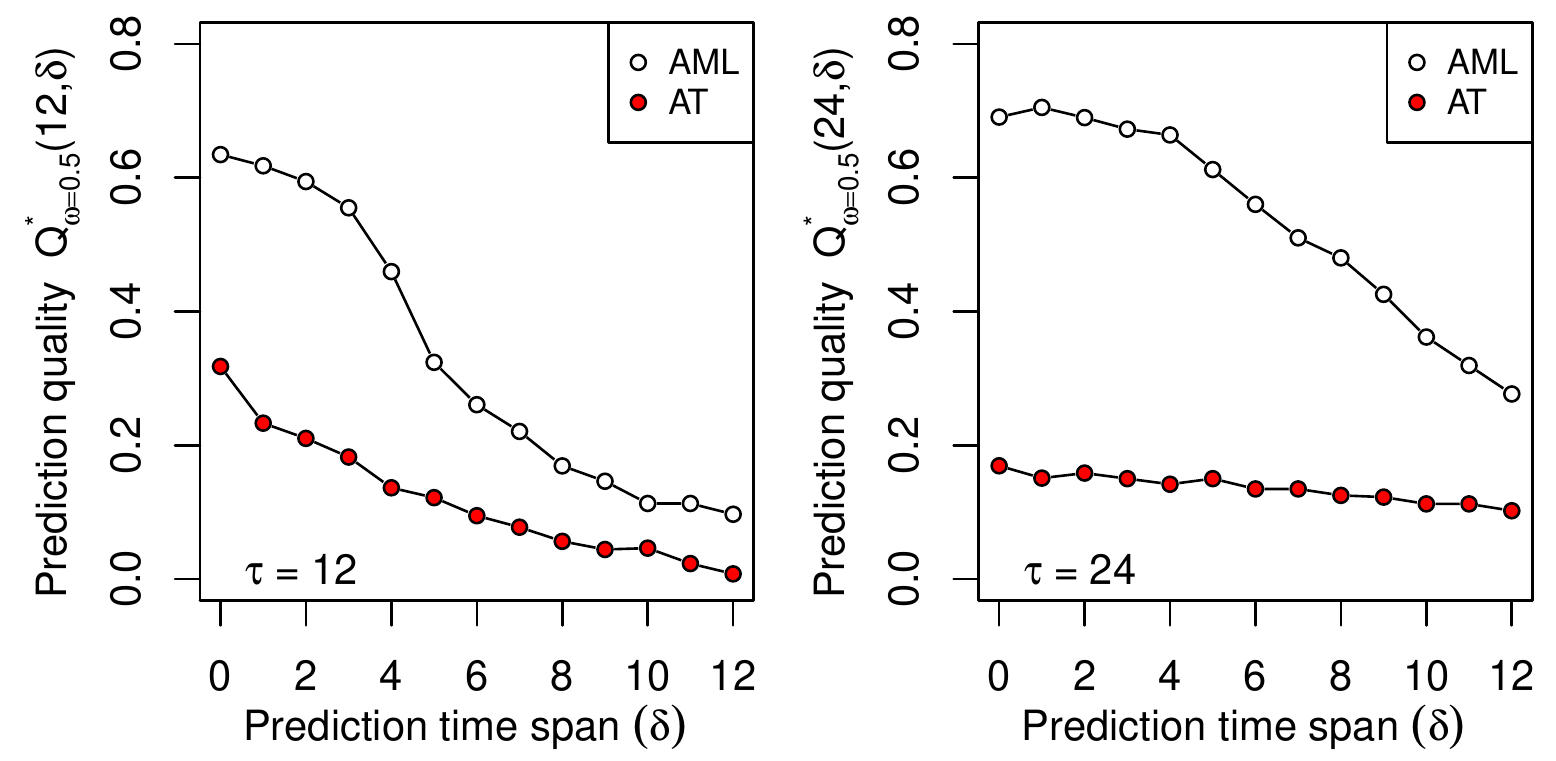}
    \caption{The prediction qualities of the AML and AT model at some horizons.}
    \label{fig:predict:quality}
\end{figure}

\remark*[Fabio]{A paragraph explaining why we need the predictability below.}

We should assess the predictability of a VDM not only along the prediction time
span, but also along the horizon to ensure that the VDM is able to consistently
predict the vulnerability data in an expected period. To facilitate such
assessment we introduce the \emph{predictability} metric which is the average
of prediction qualities at a certain horizon.

The predictability of the curve \curve{\odp} at the horizon \horizon\ in a time
span of $\Timespan$ months is defined as the average of the prediction quality
of \curve{\odp} at the horizon \horizon\ and its $\Timespan$ consecutive
horizons $\horizon + 1, \horizon+2,..., \horizon+\Timespan$, as the following
equation shows:
\begin{align}
    \predict(\horizon, \Timespan) &= \sqrt[\Timespan+1]{\prod_{\timespan=0}^\Timespan
    Q^*_\omega(\horizon,\timespan)} \label{eq:predict}
\end{align}
where \Timespan\ is the prediction time span.

In \eqref{eq:predict}, we use the geometric mean instead of the arithmetic
mean. The temporal quality is a normalized measure so using the arithmetic mean
to average such values might produce a meaningless result, whereas the
geometric mean behaves correctly \cite{FLEM-WALL-86-CACM}.

\subsection{\ref{step:comparison}: Compare VDM}
This section addresses the second research question \ref{rq:comparison}. The
comparison is based on the quality and the predictability of VDMs. The base
line for the comparison is that: \emph{the better model is a better one in
forecasting changes}. Hereafter, we discuss how to compare VDM:

Given two models \vdm[1] and \vdm[2], the comparison between \vdm[1] and
\vdm[2] could be done as below:

We compare the predictability of \vdm[1] and that of \vdm[2]. Let $\rho_1,
\rho_2$ be the predictability of \vdm[1] and \vdm[2], respectively.
\begin{equation}\label{eq:compare:predict}
\begin{aligned}
    \rho_1 &= \Set{\predict[\omega=0.5](\horizon, \Timespan)|\horizon = 6..\horizon_{max}, \vdm[1]} \\
    \rho_2 &= \Set{\predict[\omega=0.5](\horizon, \Timespan)|\horizon = 6..\horizon_{max}, \vdm[2]}
\end{aligned}
\end{equation}
where the prediction time span \Timespan\ could follow the criteria
\ref{cr:timespan}; $\horizon_{max} = \min(72,  \max_{\release \in
\Release}\horizon^\release_{max})$. We employ the one-sided Wilcoxon rank-sum
test to compare $\rho_1, \rho_2$. If the returned \pvalue\ is less than the
significance level $\alpha = 0.05$, the predictability of \vdm[1] is
stochastically greater than that of \vdm[2]. It also means that \vdm[1] is
better than \vdm[2]. If $\pvalue \ge 1-\alpha$, we conclude the opposite \ie
\vdm[2] is better than \vdm[1]. Otherwise we have not enough evidence either
way.

If the previous comparison is inconclusive, we retry the comparison using the
value of temporal quality of the VDMs instead of the predictability. We just
replace $Q_{\omega=0.5}(\horizon)$ for \predict[\omega=0.5](\horizon,
\Timespan) in the equation \eqref{eq:compare:predict}, and repeat the above
activities.

When we compare models \ie we run several hypothesis tests, we should pay
attention on the familywise error rate which is the probability of making one
or more type I errors. To avoid such problem, we should apply an appropriate
controlling procedure such as the Bonferroni correction. In the case above, the
significance level by which we conclude a model is better than another one is
divided by the number of tests performed.

\begin{example}
    When we compare one model against other seven models, the Bonferroni-corrected
    significance level is: $\alpha = ^{0.05}/_7 \approx 0.007$.
\end{example}

The above comparison activities are summarized in the criteria \ref{cr:compare}
(see \tabref{tbl:methodology}).


\section{An Assessment on Existing VDMs}\label{sec:exp}
We apply the above methodology to assess the performance of eight existing VDMs
(see also \tabref{tbl:vdm}). The experiment evaluates these VDMs on $30$
releases of the four popular web browsers: IE, Firefox, Chrome, and Safari.
Here, only the formulae of these models are provided. More detail discussion
about these models as well as the meaning of their parameters are referred to
their corresponding original work.

\begin{table}
    \centering
    \caption{The VDMs in evaluation and their equation.}
    \extcaption[1\columnwidth]{This table presents the list of VDMs and their equation in the alphabetical order.
    The meaning of each parameter should be found in original work of each model. }
    \label{tbl:vdm}
    \scriptsize
    \resizebox{\columnwidth}{!}{
    \begin{tabular}{l>{$\displaystyle}l<{$}}
      \toprule
      Model & Equation  \\
      \midrule
      Alhazmi-Malaiya Logistic (AML) & \Omega(t) = \frac{B}{BCe^{-ABt} + 1}  \\[5pt]
      Anderson Thermodynamic (AT) & \Omega(t) = \frac{k}{\gamma}\ln(t) + C   \\
      Joh Weibull (JW) & \Omega(t) = \gamma(1 - e^{-\Parenthese{\frac{t}{\beta}}^\alpha}) \\[2pt]
      Linear (LN) & \Omega(t) = At + B  \\ [2pt]
      Logistic Poisson (LP) & \Omega(t) = \beta_0\ln(1 + \beta_1t)  \\ [2pt]
      Rescorla Exponential (RE) & \Omega(t) = N(1 - e^{-\lambda t})  \\
      Rescorla Quadratic (RQ) & \Omega(t) = \frac{At^2}{2} + Bt  \\[5pt]
      Younis Folded (YF) & \Omega(t) = \frac{\gamma}{2} \left[
                    \operatorname{erf}\Parenthese{\frac{t-\tau}{\sqrt{2\sigma}}}
                     +
                     \operatorname{erf}\Parenthese{\frac{t+\tau}{\sqrt{2\sigma}}}\right]
                      \\
      \bottomrule
    \end{tabular}}
    \extcaption{\scriptsize \emph{Note}: \emph{erf()} is the error function, $\displaystyle\operatorname{erf}(x) =
    \frac{2}{\sqrt{\pi}}\int^x_0 e^{-t^2}\;dt$}
    \ACCORCIA[-1]
\end{table}

\begin{table}
    \centering
    \caption{Vulnerability data sources of browsers.}
    \label{tbl:datasources}
    \resizebox{\columnwidth}{!}{\begin{tabular}{lcc}
        \toprule
        Data Source & Category & Apply for \\
        \midrule
        National Vulnerability Database (NVD) & \ds{TADV} & All browsers \\
        Mozilla Foundation Security Advisory (MFSA) & \ds{ADV} & Firefox \\
        Mozilla Bugzilla (MBug) & \ds{BUG} & Firefox \\
        Microsoft Security Bulletin (MSB) & \ds{ADV} & IE \\
        Apple Knowledge Base (AKB) & \ds{ADV} & Safari \\
        Chrome Issue Tracker (CIT) & \ds{BUG} & Chrome \\
        \bottomrule
    \end{tabular}}
\end{table}

\begin{table}
    \caption{Collected data sets.}
    \extcaption[1\columnwidth]{Bullets (\tick) indicate available data sets. Dashes (---) mean there
    is no data sources available to collect the data sets.}
    \label{tbl:datasets}
    \scriptsize
    \resizebox{\columnwidth}{!}{
    \begin{tabular}{l*{6}{@{\hspace{2pt}}c}C{5ex}}
        \toprule
         & Releases & \ds{NVD} & \ds{NVD.Bug} & \ds{NVD.Advice} & \ds{NVD.NBug} & \ds{Advice.Nbug} & Total
         Datasets \\
        \midrule
        Chrome & 12(\ver{1.0}--\ver{12.0}) & \tick & \tick & --- & \tick & --- & 36 \\
        Firefox & 8(\ver{1.0}--\ver{5.0}) & \tick & \tick & \tick & \tick & \tick & 40  \\
        IE  & 5(\ver{4.0}--\ver{8.0}) & \tick & --- & \tick & --- & --- & 10 \\
        Safari & 5(\ver{1.0}--\ver{5.0}) & \tick & --- & \tick & --- & --- & 10 \\
        \midrule
        \textbf{Total} & 30 &  &  &  & & & 96\\
        \bottomrule
    \end{tabular}}
\end{table}

\subsection{Data Acquisition}

\tabref{tbl:datasources} presents the availability of vulnerability data
sources for the browsers in our study. For each data source, the table reports
the name, the category (see also \secref{sec:step:data}), and the browser that
the data source maintains vulnerability data. We use \ds{NVD} as a
representative third-party data source due to its popularity in past studies.
This makes our work comparable with previous ones.

\tabref{tbl:datasets} reports data sets collected for this experiment (see also
\tabref{tbl:datasets:definition} for the classification). In total, we
collected 96 data sets for 30 major releases. In the table, we use the bullet
(\tick) to indicate the availability of data sets. In these collected data
sets, we extracted a total of $4,063$ observed samples.

\subsection{The Applicability of VDMs}\label{sec:exp:applicability}
We ran model fitting algorithms for these observed samples by using R
\ver{2.13}. Model fitting took about $82$ minutes on a dual-core 2.73GHz
Windows machine with 6GB of RAM yielding $32,504$ curves in total.

\subsubsection{Goodness-of-Fit Analysis for VDMs}

\begin{figure*}
    \centering
    \includegraphics[width=0.95\textwidth, height=17\baselineskip]{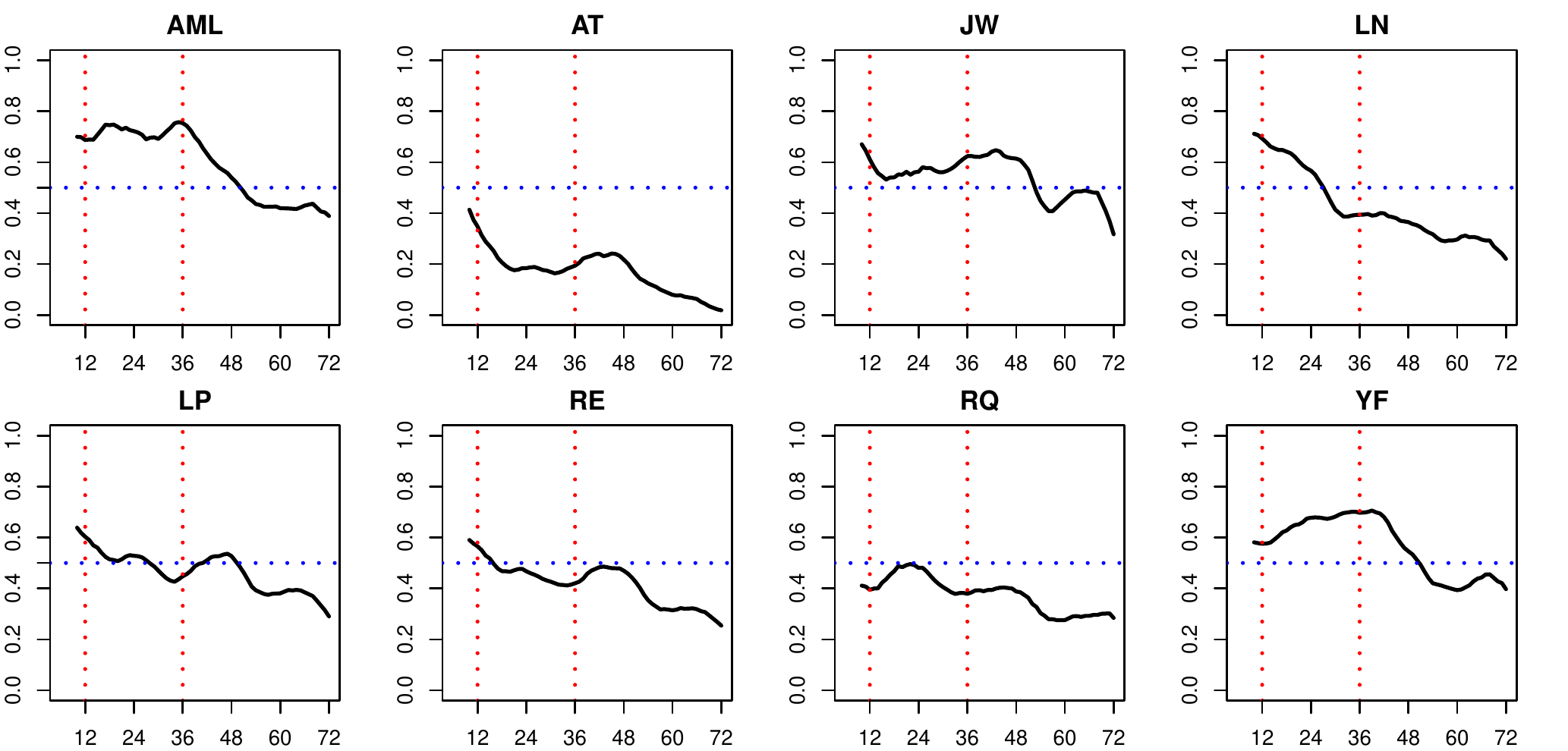}
    \extcaption[\textwidth]{The X-axis is the number of months since release (\ie horizon \horizon). The Y-axis
    is the value of temporal quality. The solid lines are the moving average of $Q_{\omega=0.5}(\horizon)$ with
    window size $k=5$. The dotted horizontal line at $0.5$ is the base line to assess VDM. Vertical lines
    are the marks of the horizons of $12^{th}$ and $36^{th}$ month. }
    \caption{The trend of temporal quality $Q_{\omega=0.5}(\horizon)$ of the VDMs in first 72 months.}
    \label{fig:quality:evolution}
\end{figure*}

\tabref{tbl:gof:lasthorizon} reports the goodness-of-fit of existing VDMs on
the largest horizons of browser releases, using the \ds{NVD} data sets. In
other words, we use following observed samples to evaluate all models:
\[
    \DP_{\ds{NVD}} = \Set{\series{\release, \ds{NVD}, \horizon^\release_{max}}|\release \in \Release}
\]
where \Release\ is the set of all releases mentioned in \tabref{tbl:datasets}.
\tabref{tbl:gof:lasthorizon} provides a view that previous studies often used
to report the goodness-of-fit for their proposed models. To improve
readability, we report the categorized goodness-of-fit based on the \pvalue\
(see \ref{cr:gof}) instead of the raw \pvalue s. In this table, we use a check
mark (\checkmark), a blank, and a cross ($\times$) to respectively indicate a
\gfit, an \ifit, and a \nfit. Cells are shaded accordingly to improve the
visualization effect. The table shows that two models AT and RQ have a very
high ratio of \nfit\ ($0.9$ and $0.7$, respectively); whereas, all other models
have their ratio of \nfit\ less than \Qthreshold. We should observe that this
is a very large time interval and some systems have long gone into retirement.
For example, FF \ver{2.0} vulnerabilities are no longer sought by researchers.
They are a byproduct of research on later versions.

To have a more realistic picture, we also study the temporal quality. The
inconclusiveness contribution factor $\omega$ is set to $0.5$ as described in
\ref{cr:quality}. \figref{fig:quality:evolution} exhibits the moving average
(windows size $k=5$) of the $Q_{\omega}(\horizon)$. The dotted vertical lines
marks horizon $12$ (when software is young), and $36$ (when software is
middle-age). We cut the temporal quality at horizon $72$ though we have more
data for some systems (\eg IE \ver{4}, FF \ver{1.0}). It is because that after
6 years software is very old, the vulnerability data reported for such releases
might be not reliable, and might overfit the VDMs. The dotted horizon line at
$0.5$ is used as a base line to assess VDMs.


Clearly from the temporal quality trends in \figref{fig:quality:evolution} both
AT and RQ models should be rejected since their temporal quality always sinks
below the base line. Other models may be adequate when software is young
(before 12 months). The AML and LN models look better than other models in this
respect.

When software is middle-age (between 12 and 36 months), the AML model is still
relatively good. JW and YF improve when approaching month $36^{th}$ though JW
get worse after month $12^{th}$. The quality of both LN and LP worsen after
month $12^{th}$, and sink below the base line when approaching month $36^{th}$.
RE is almost below the base line after month $15^{th}$. Hence, in the
middle-age period, AML, JW, and YF models may turn to be adequate; LN and LP
are deteriorating but might be still considered adequate; whereas RE should
clearly be rejected.


When software is old (36+ months), AML, JW, and YF deteriorate and go below the
base line at month $48^{th}$ (approximately); other models also collapse below
the base line.

\figref{fig:quality:distribution} summarizes the distribution of VDM temporal
quality in three period: software is young (before 12 moths), software is
middle-age (13 to 36 months), and software is old (37 to 72 months). The red
horizonal line at $0.5$ is the base line. We additionally colour these box
plots according to the comparison between the corresponding distribution and
the base line as follows:
\begin{itemize}
    \item white: the distribution is significantly greater than the base
        line;
    \item dark gray: the distribution is significantly less than the base
        line (we should reject the models outright);
    \item gray: the distribution is not statistically different from the
        base line.
\end{itemize}
The box plots clearly confirm our observation in
\figref{fig:quality:evolution}.  Both AT and RE models are all significantly
below the base line.  AML, JW, and YF modes are significantly above the base
line when software is young and middle age, and not statistically different
from the base line when software is old. LN and LP models are significantly
greater than the base line when software is young, but they deteriorate for
middle-age software, and significantly collapse below the base line for old
software.

In summary, our quality analysis shows that:
\begin{itemize}
    \item AT and RQ models should be rejected.
    \item All other models may be adequate when software is young. Only
        s-shape models (\ie AML, YW, YF) might be adequate when software is
        middle-age.
    \item No model is good when the software is too old.
\end{itemize}

\tableNVDGoF[A potentially misleading results of overfitting VDMs in the
largest horizon of browser releases, using \ds{NVD} data
sets]{\label{tbl:gof:lasthorizon}}

\begin{figure}
    \centering
    \includegraphics[width=1.05\columnwidth, height=10\baselineskip]{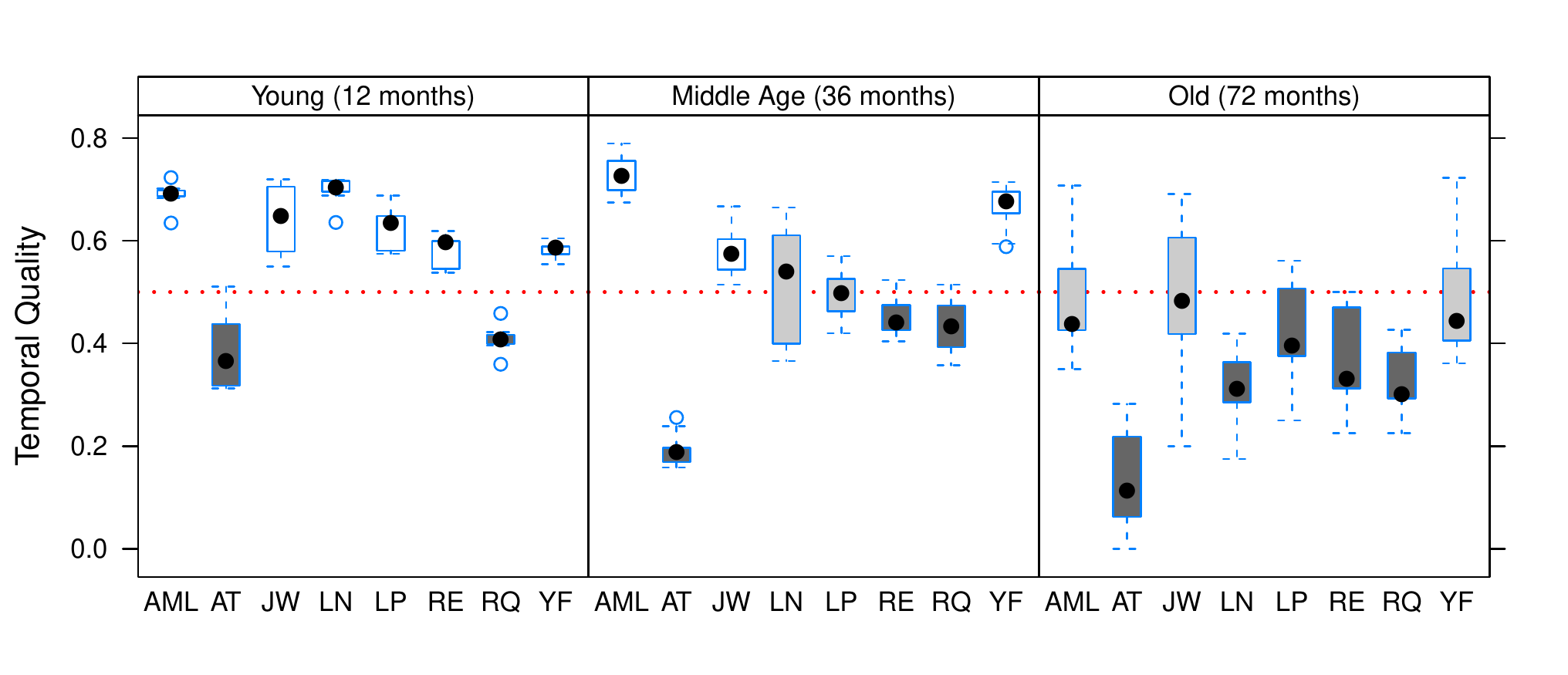}
    \ACCORCIA[-2]
    \extcaption[\columnwidth]{A horizonal line at value of $0.5$ is used as the base line to justify temporal quality.
    Box plots are coloured with respect to the comparison between the corresponding distribution and the base
    line:  white - significantly above the base line, gray - no statistical difference, dark gray - significantly
    below the base line (\ie rejected).}
    \caption{The temporal quality distribution of each VDM in different periods of software lifetime.}
    \label{fig:quality:distribution}
\end{figure}





\subsubsection{Predictability Analysis for VDMs}

From the previous quality analysis, AT and RQ models are low quality, and they
should not be considered for all periods of software lifetime. Hence, we
exclude these models from the predictability analysis. Furthermore, since no
model is good when software is too old, we analyze the predictability of these
models only for the first $36$ months since the release of a software. This
period is still a large time if we consider that most recent releases live less
than a year.

\figref{fig:predict} reports the moving average (windows size equals $5$) for
the trends of VDMs' predictability along horizons in different prediction time
spans. The dotted horizonal line at value of $0.5$ is the base line to assess
the predictability of VDMs (as same as the temporal quality of VDMs).

\begin{figure}
    \centering
    \includegraphics[width=1.02\columnwidth]{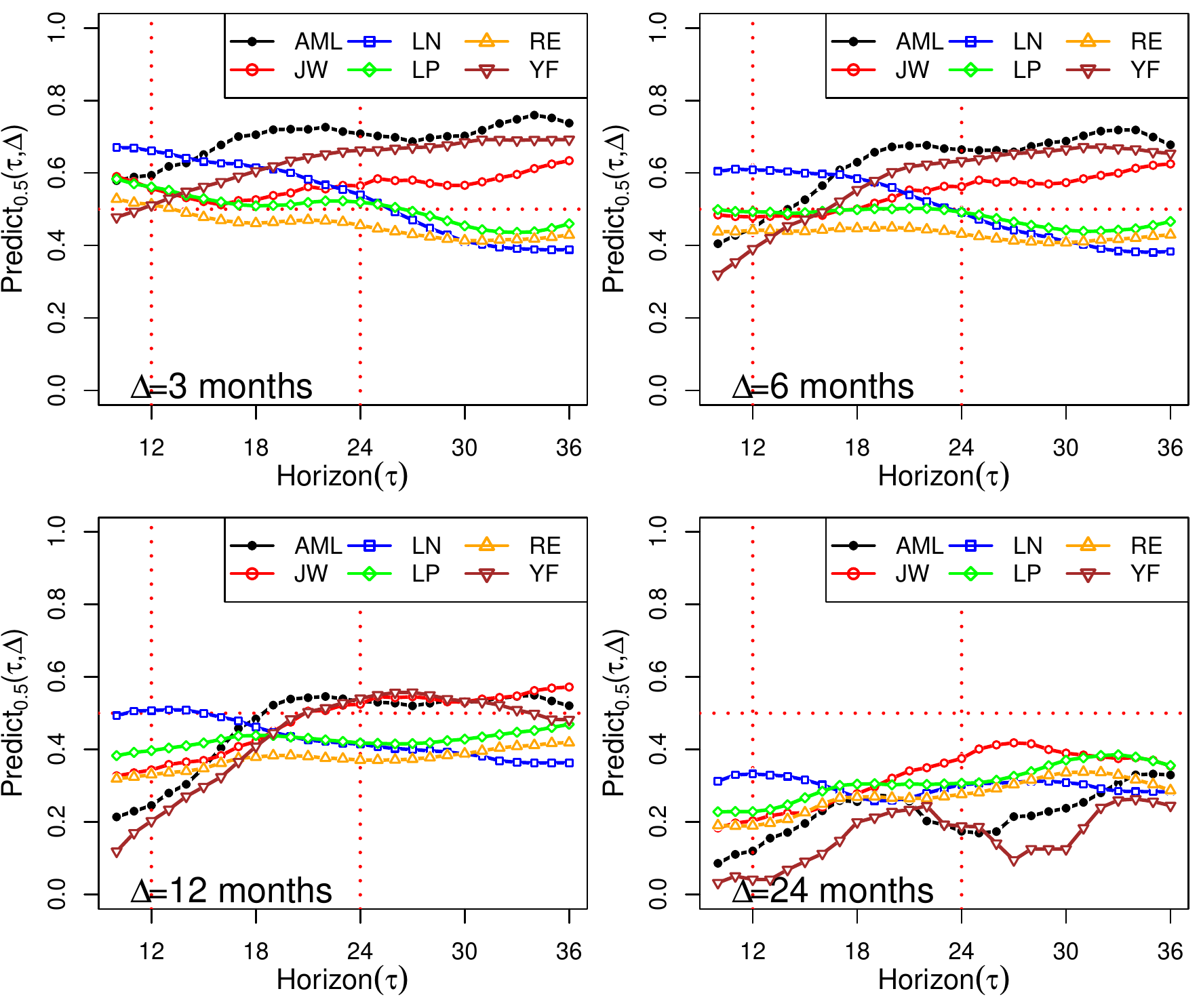}
    \ACCORCIA[-1.5]
    \extcaption[\columnwidth]{A horizonal line at value of $0.5$ is the base line to assess the predictability.}
    \caption{The predictability of VDM in different prediction time spans (\Timespan).}
    \label{fig:predict}
    \ACCORCIA
\end{figure}

When the prediction time span is short ($3$ months), the predictability of LN,
AML, JW, and LP models is above the base line for young software ($12$ months).
When software is approaching month $24^{th}$, though decreasing the
predictability of LN is still above the base line, but goes below the base line
after month $24^{th}$. The LP model is no different with the base line before
month $24^{th}$, but then also goes below the base line. In contrast, the
predictability of AML, YF and JW are improving with age. They are all above the
base line until the end of the study period (month $36^{th}$. Therefore, only
s-shape models (AML, YF, and JW) may be adequate for middle-age software.

For the medium prediction time span of 6 months, only the LN model may be
adequate (above the base line) when software is young, but becomes inadequate
(below the base line) after month $24^{th}$. In the meanwhile S-shape models
are inadequate for young software, but are improving quickly later. They may be
all adequate after month $18^{th}$ and keep this performance until the end of
the study period.

When the prediction time span is long (\ie 12 months), all models (except LN)
sink below the base line for young software. The LN model is not significantly
different from the base line. In other words, no model could be adequate for
young software in this prediction time span. After month $18^{th}$, the AML
model goes above the base line, and after month $24^{th}$, all s-shape models
are above the base line. Hence they may be all adequate. Their performances are
somewhat unchanged for the remain period.

When the prediction time span is very long (\ie 24 months) no model is good
enough as all models sink below the base line.

In summary, our predictability analysis shows that:
\begin{itemize}
    \item For a short prediction time span (\ie 3 months), the
        predictability of LN, AML, and LP models may be adequate for young
        software. Hence they could be considered for the scenario
        \emph{Plan for short-term support}. When software is approaching
        middle-age, s-shape models (AML, JW, YF) are better than others.
    \item For a medium (\ie 6 months) and long (\ie 12 month) prediction
        time spans, only the predictability of the LN model may be adequate
        for young software. And therefore this model could be appropriate
        the purpose of the scenarios \emph{Upgrade of keep} and \emph{Plan
        for long-term support}. When software is approaching middle-age,
        only s-shape models (AML, JW, YF) may be adequate and might be
        considered for planning the long-term support and for studying
        historical trends (\ie scenario \emph{Historic analysis}).
    \item For a very long prediction time span (\ie 24 months), no model
        has a good enough predictability.
\end{itemize}

\subsection{Comparison of Existing VDMs}
The comparison between VDMs follows \ref{step:comparison}. Instead of reporting
tables of \pvalue s, we visualize the comparison result in terms of directed
graphs where nodes represent models, and connections represent the order
relationship between models.

\figref{fig:comparison} summarizes the comparison results between models in
different settings of horizons (\horizon) and prediction time spans
(\Timespan). A directed connection from two models determines that the source
model is better than the target model in terms of either predictability, or
quality, or both. The line style of the connection depended on the following
rules:
\begin{itemize}
    \item \emph{Solid line}: the predictability and quality of the source
        is significantly better than the target's.
    \item \emph{Dashed line}: the predictability of the source is
        significantly better than the target.
    \item \emph{Dotted line}: the quality of the source is significantly
        better than the target.
\end{itemize}

By term \emph{significantly}, we means the \pvalue\ of the corresponding
one-side Wilcoxon rank-sum test is less than the significance level. We apply
the Bonferroni correction to control the multi comparison problem, hence the
significance level is: $\alpha = 0.05/5 = 0.01$.

\begin{figure}
    \def\graphwidth{0.455\columnwidth}
    \centering
    \subfigure[Young software, short term prediction ($\horizon=6..12, \Delta=3$)] {
        \includegraphics[width=\graphwidth]{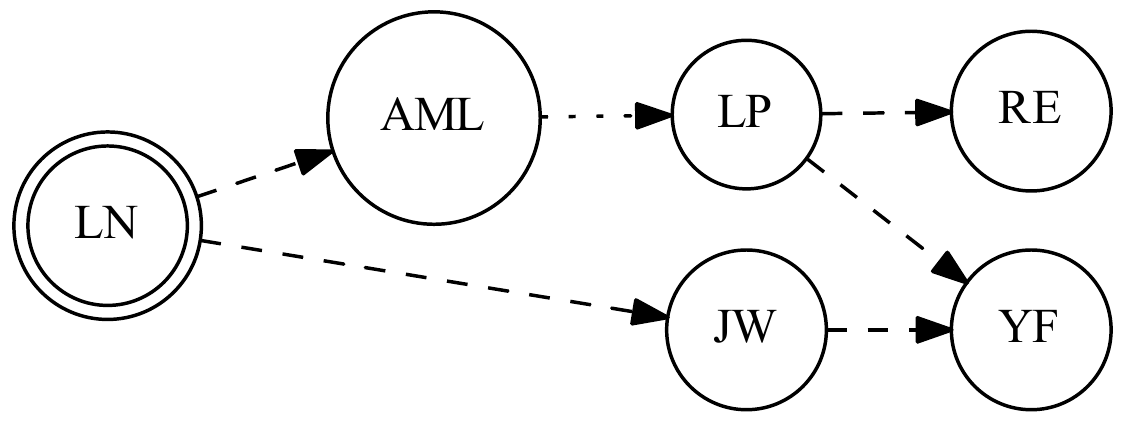}
        \label{fig:comparison:all}
        \ACCORCIA[-2]
    }
    \subfigure[Young software, medium term prediction ($\horizon=6..12, \Delta=6$)] {
        \includegraphics[width=\graphwidth]{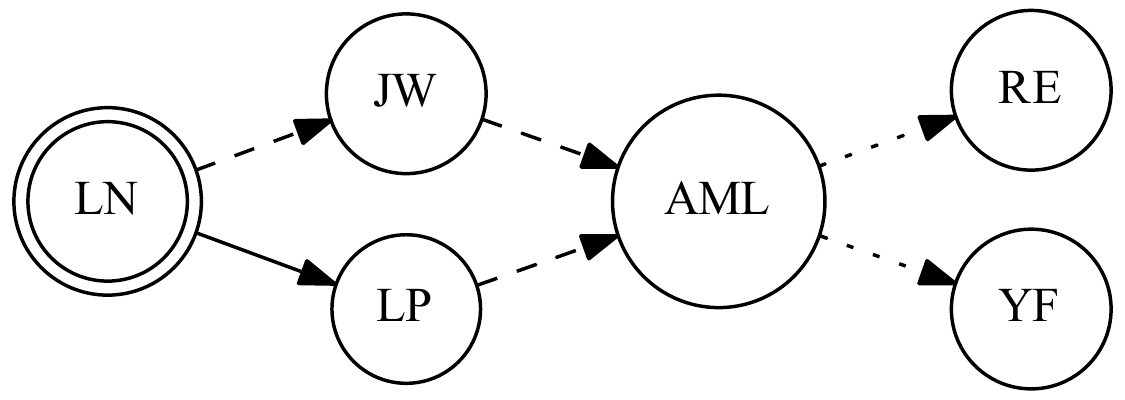}
        \label{fig:comparison:36}
    }
    \subfigure[Young software, 1 year prediction ($\horizon=6..12, \Delta=12$)] {
        \includegraphics[width=\graphwidth]{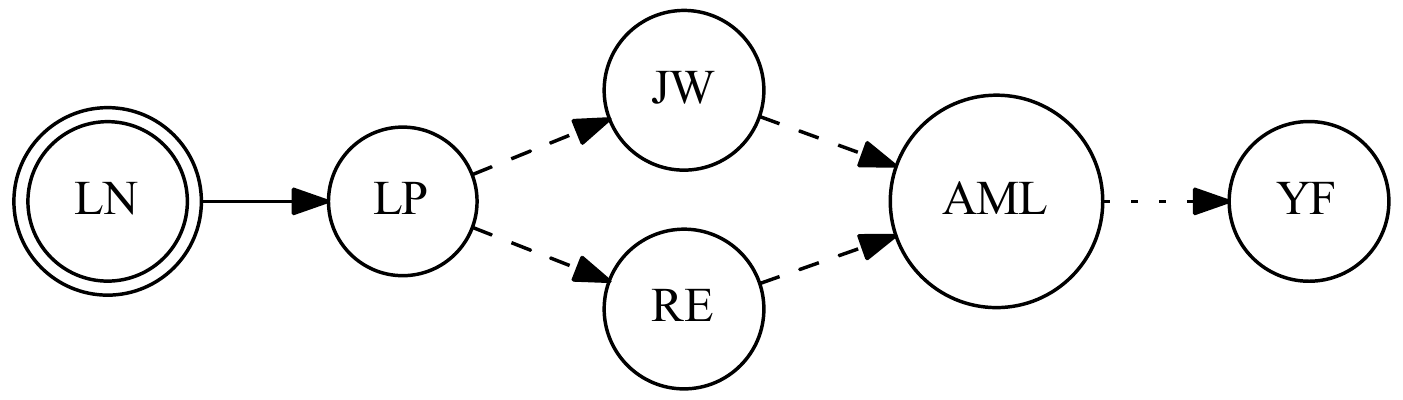}
        \label{fig:comparison:30}
        \ACCORCIA[-2]
    }
    \subfigure[Middle-age software, short term prediction ($\horizon=12..24, \Delta=03$)] {
        \includegraphics[width=\graphwidth]{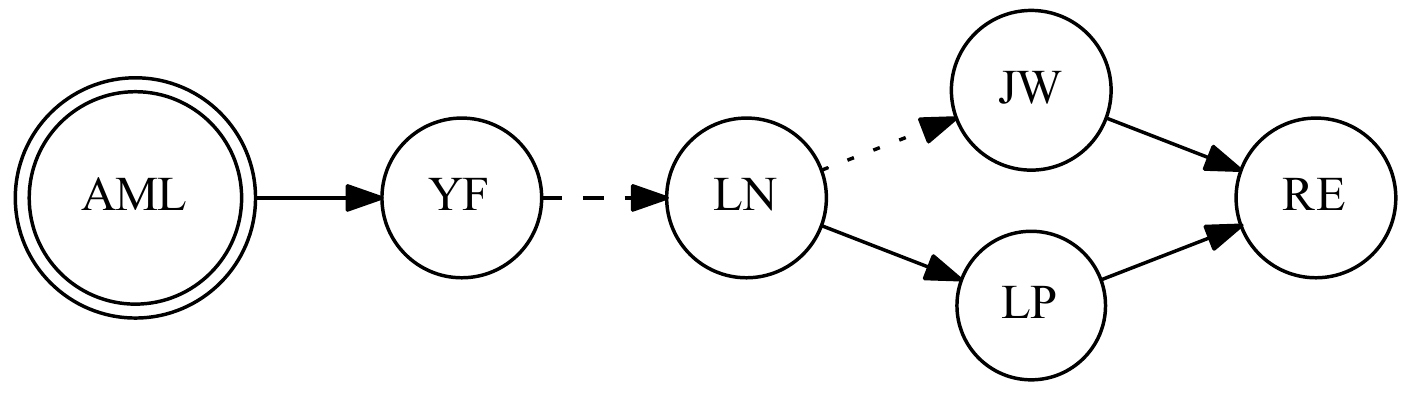}
        \label{fig:comparison:30}
    }
    \subfigure[hang,centerlast][Middle-age software, 1 year prediction ($\horizon=12..24, \Delta=12$)] {
        \includegraphics[width=0.4\columnwidth]{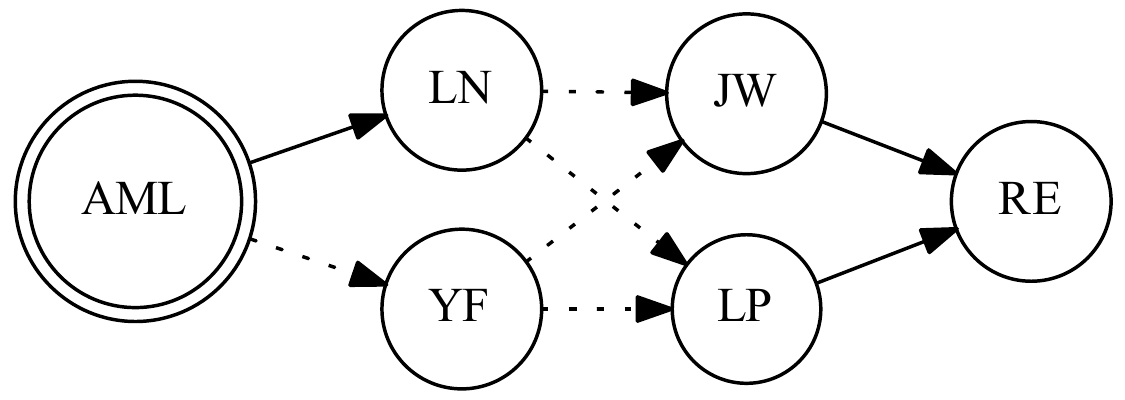}
        \label{fig:comparison:30}
    }
    \subfigure[hand,centerlast][Old software, 1 year prediction \newline($\horizon=24..36, \Delta=12$)] {
        \includegraphics[width=0.52\columnwidth]{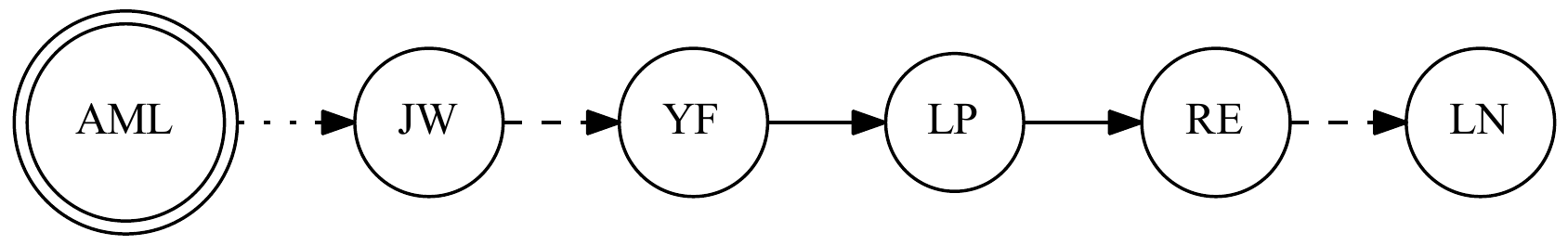}
        \label{fig:comparison:30}
    }
    \extcaption[\columnwidth]{A directed connection from two nodes determines that the source
    model is better than the target one with respect to their predictability (dashed line), or their
    quality (dotted line), or both (solid line). A double cirlce marks the best model.
    RQ and AT are not shown as they are the worst models.}
    \caption{The comparison results among VDMs in some usage scenarios.}
    \label{fig:comparison}
\end{figure}

\begin{table}
    \centering
    \caption{Suggested models for different usage scenarios.}
    \label{tbl:comparison:result}
    \resizebox{1\columnwidth}{!}{\begin{tabular}{lccc}
        \toprule
         & Observation & Prediction \\
        Scenario & Period & Time Span & Model(s) \\
        \midrule
        Plan for short-term support & 6--12 & 3 & LN  \\
         & 12--24 & 3 & AML  \\
        Plan for long-term support & 6--12 & 12 & LN \\
        & 12--24 & 12 & AML \\
        Upgrade or keep & 6--12 & 6 & LN \\
        Historic analysis & 24--36 & 12 & AML \\
        \bottomrule
    \end{tabular}}
    \extcaption{Note: the unit is month.}
    \ACCORCIA
\end{table}

According to the figure, \tabref{tbl:comparison:result} suggests model(s) for
different usage scenario described in the criteria \ref{cr:timespan} (see
\tabref{tbl:methodology}).

In short, \emph{when software is young, the LN model is the most appropriate
choice. This is because the vulnerability discovery is linear. When software is
approaching middle-age, the AML model becomes superior}.

\section{Threats to Validity}\label{sec:validity}
{\bf Construct validity} includes threats affecting the way we collect
vulnerability data and the way we generate VDM curves with respect to the
collected data. Following threats in this category are identified:
\begin{itemize}
    \item {\bf Bugs in data collector.} Most of vulnerability data are
        available in HTML pages. We have developed a web crawler to extract
        interesting feature from HTML page, and also XML data. The employed
        technique is as same as the one discussed in
        \cite{MASS-NGUY-10-METRISEC}. The crawler might be buggy and could
        generate errors in data collection. To minimize such impact, we
        have tested the crawler many times before collecting the data. Then
        by randomly checking the collected data, when an error is found we
        corrected the corresponding bug in the crawler and recollected the
        data.
    \item {\bf Bias in \emph{bug-to-nvd} linking scheme.} While collecting
        data for \ds{Advice.Nbug}, we apply some heuristic rules to link a
        \ds{bug} to an \ds{NVD} entry based on the relative position in the
        MFSA report. We manually checked many links for the relevant
        connection between bug reports and NVD entries. All checked links
        were found to be consistent. Some errors might still creep in this
        case.
    \item {\bf Bias in \emph{bug-affects-version} identification.} We do
        not have a complete assurance that a security bug affects to which
        versions. Consequently, we assume that a bug affects all versions
        mentioned in the linked \ds{NVD}. This might overestimate the
        number of bugs in each version. To mitigate the problem, we
        estimate the latest release that a bug might impact, and filter all
        vulnerable releases after this latest. Such estimation is done
        thank to the mining technique discussed in \cite{SLIW-05-MSR}. We
        further discuss these types of errors in NVD in
        \cite{NGUY-MASS-13-ASIACCS}. These errors only affect the fitness
        of models over the long term so only valuations after the $24$ or
        $36$ months might be affected.
    \item {\bf Error in curve fitting.} From the collected vulnerability,
        we estimate the parameters of VDMs by using the Nonlinear
        Least-Square technique implemented in R (\texttt{nls()} function).
        This might not produce the most optimal solution and may impact the
        goodness-of-fit of VDMs. To mitigate this issue, we additionally
        employed a commercial tool \ie CurveExpert
        Pro\footnote{\url{http://www.curveexpert.net/}, site visited on 16
        Sep, 2011} to cross check the goodness-of-fit in many cases. The
        results have shown that there is no difference between R and
        CurveExpert.
\end{itemize}

\ACCORCIA[1] \noindent {\bf Internal validity} concerns the causal relationship
between the collected data and the conclusion drawn in our study. Here, we have
identified the following threats that might bias our conclusion.
\begin{itemize}
    \item {\bf Bias in statistics tests.} Our conclusions are based on
        statistics tests. These tests have their own assumptions. Choosing
        tests whose assumptions are violated might end up with wrong
        conclusions. To reduce the risk we carefully analyzed the
        assumptions of the tests to make sure no unwarranted assumption was
        present. We did not apply any tests with normality assumptions
        since the distribution of vulnerabilities is not normal.
\end{itemize}

\ACCORCIA[1] \noindent {\bf External validity} is the extent to which our
conclusion could be generalized to other scenarios. Our experiment is based on
the vulnerability data of some major releases of the four most popular browsers
covering almost all market shares. Therefore we can be quite confident about
our conclusion for browsers in general. However, it does not mean that our
conclusion is valid for other types of application such as operating systems.
Such validity requires extra experiments.

\section{Related Work}\label{sec:relatedwork}
Anderson \cite{ANDE-02-OSS} proposed a VDM (a.k.a. Anderson Thermodynamic, AT)
based on reliability growth models, in which the probability of a security
failure at time $t$, when $n$ bugs have been removed, is in inverse ratio to
$t$ for alpha testers. This probability is even harder for beta testers,
$\lambda$ times more than alpha testers. However, he did not conduct any
experiment to validate the proposed model. Our results show that this model is
not appropriate. This is a first evidence that reliability and security obey
different laws.

Rescorla \cite{RESC-05-SP} also proposed two mathematical models, called
\emph{Linear model} (a.k.a Rescorla Quadratic, RQ) and \emph{Exponential model}
(a.k.a Rescorla Exponential, RE). He has performed an experiment on four
versions of different operation systems (\ie Windows NT 4.0, Solaris 2.5.1,
FreeBSD 4.0 and RedHat 7.0). In all cases, the goodness-of-fit of these two
models were inconclusive since their \pvalue\ ranged from $0.167$ to $0.589$.
Rescorla discussed many shortcomings of NVD, but his study heavily relied on it
nonetheless.

Alhazmi and Malaiya \cite{ALHA-MALA-05-ISSRE} proposed another VDM inspired by
s-shape logistic model, called \emph{Alhazmi Malaiya Logistic} (AML). The
intuition behind the model is to divide the discovery process into three
phases: \emph{learning phase, linear phase} and \emph{saturation phase}. In the
first phase, people need some time to study the software, so less
vulnerabilities are discovered. In the second phase, when people get deeper
knowledge of the software, much more vulnerabilities are found. In the final
phase, since the software is out of date, people may lose interest in finding
new vulnerabilities. So cumulative vulnerabilities tend to stable. In
\cite{ALHA-MALA-05-ISSRE}, the authors validated their proposal against several
versions of Windows (\ie Win 95/98/NT4.0/2K) and Linux (\ie RedHat Linux 6.1,
7.1). Their model fitted Win 95 very well (\emph{p-value} $\approx 1$), and Win
NT4.0 (\emph{p-value} = $0.923$). For other versions, their own validation
showed that the AML model was inconclusive (\ie the \emph{p-value} ranged from
$0.054$ to $0.317$).

In another work, Alhazmi and Malaiya \cite{ALHA-MALA-08-TR} compared their
proposed model with Rescorla's \cite{RESC-05-SP} (RE, RQ) and Anderson's
\cite{ANDE-02-OSS} (AT) on Windows 95/XP and Linux RedHat Linux 6.2, Fedora.
The result shows that their logistic model has a better goodness-of-fit than
others. For Windows 95 and Linux 6.2, as the vulnerabilities distribute along
s-shape-like curves, only AML is able to fit it (\emph{p-value}=1), whereas all
other models fail to match the data (\emph{p-value} $\le 0.05$). For Windows
XP, the story is different. RQ turns to be the best one with
\emph{p-value}$=0.97$, while AML poorly match the data
(\emph{p-value}=$0.147$).

Woo \etal \cite{WOO-ALHAZMI-MALAIYA-06-SEA} carried out an experiment with AML
model on three browsers IE, Firefox and Mozilla. However, it is unclear which
versions of these browsers were analyzed. Most likely, they did not distinguish
between versions. As discussed in section \secref{sec:questions} (\eg
\ref{ex:bias:multiversions}), this could largely bias their final result. In
their experiment, IE has not been fitted, Firefox was fairly fitted, and
Mozilla was good fitted. From this result, we could not conclude any thing
about the performance of AML. In another experiment, Woo \etal
\cite{WOO-etal-11-CS} validated AML against two web servers: Apache and IIS.
Also, they did not distinguish between versions of Apache and IIS. In this
experiment, AML has demonstrated a very good performance on vulnerability data
(\emph{p-value} $=1$).

Kim \etal \cite{KIM-etal-07-HASE} introduced the Multiple-Version Discovery
Model (MVDM) which is the generalization of AML. The MVDM separated the
cumulative vulnerabilities of a version into several fragments where the first
fragment captured the vulnerabilities affecting this version and past versions,
and the other fragments are the shared vulnerabilities of this version and
future versions. The MVDM basically is the weighted aggregation of individual
AML model in these fragments. The weights are determined by the ratios of
shared code between this version and future ones. The goodness-of-fit of MVDM
has been compared with AML in two versions of Apache and two version of MySQL.
As the result, both AML and MVDM were well fitted against the data ($\pvalue
\ge 0.99$). MVDM might be better but the difference was quite negligible.

Joh \etal\cite{JOH-etal-08-ISSRE} proposed a VDM based on the Weibull
distribution. The proposed model was also compared with the AML model in two
versions of Windows (XP, Server 2007) and two versions of Linux (RedHat Linux
and RedHat Enterprise Linux). In that evaluation, the goodness-of-fit of the
proposed model was compared with the AML model.

Younis \etal\cite{YOUNIS-etal-11-SAM} exploited the Folded distribution to
model the discovery of vulnerabilities. The authors also compared the proposed
model with the AML model in different types of application (Windows 7, OSX 5.0,
Apache 2.0.x, and IE8). The reported results showed that the new model is
better than the AML in the cases when the learning phase is not present.


\section{Conclusion}\label{sec:conclusion}
Vulnerability discovery models have the potential to help us in predicting
future vulnerability trends. Such predictions could help individuals and
companies to adapt their software upgrade and patching schedule. However, we
have not seen any method to systematically assess these models. Hence, in this
work we have proposed an empirical methodology for VDM validation. The
methodology is built upon the analyses on the goodness-of-fit, and the
predictability of VDM at several time points during the software lifetime.
These analyses rely on two quantitative metrics: \emph{quality} and
\emph{predictability}.

We have applied this methodology to conduct an empirical experiment to assess
eight VDMs (\ie AML, AT, LN, JW, LP, RE, RQ, and YF) based on the vulnerability
data of 30 major releases of four web browsers: IE, Firefox, Chrome, and
Safari. Our experiment has revealed that:
\begin{itemize}
    \item AT and RQ models should be rejected since their quality is not
        good enough.
    \item For young software, the quality of all other models may be
        adequate. Only the predictability of LN is good enough for short
        (\ie 3 months and medium (\ie 6 months) prediction time spans,
        other models however is not good enough for latter time span.
    \item For middle-age software, only s-shape models (\ie AML, JW, and
        YF) may be adequate in terms of both quality and predictability.
    \item For old software, no model is good enough.
    \item No model is good enough for predicting results for a very long
        period (\ie 24 months in the future).
\end{itemize}


In conclusion, \emph{for young releases of browsers ($6$ -- $12$ months old) it
is better to use a linear model to estimate the vulnerabilities in the next $3$
-- $6$ months. For middle age browsers ($12$ -- $24$ months) it is better to
use an s-shape logistic model.}

In future, it is interesting to replicate our experiment in other kinds of
software, for instance operating systems and server-side applications. Based on
that, a more comprehensive assessment about the VDMs will be more solid.

\bibliographystyle{plain}

\begin{thebibliography}{10}

\bibitem{ALHA-MALA-05-ISSRE} Omar Alhazmi and Yashwant Malaiya.
\newblock Modeling the vulnerability discovery process.
\newblock In {\em Proceedings of the 16th IEEE International Symposium on
  Software Reliability Engineering (ISSRE'05)}, pages 129--138, 2005.

\bibitem{ALHA-MALA-06-ISSRE} Omar Alhazmi and Yashwant Malaiya.
\newblock Measuring and enhancing prediction capabilities of vulnerability
  discovery models for {A}pache and {IIS HTTP} servers.
\newblock In {\em Proceedings of the 17th IEEE International Symposium on
  Software Reliability Engineering (ISSRE'06)}, pages 343--352, 2006.

\bibitem{ALHA-MALA-08-TR} Omar Alhazmi and Yashwant Malaiya.
\newblock Application of vulnerability discovery models to major operating
  systems.
\newblock {\em IEEE Transactions on Reliability}, 57(1):14--22, 2008.

\bibitem{ALHA-etal-05-DAS} Omar Alhazmi, Yashwant Malaiya, and Indrajit Ray.
\newblock Security vulnerabilities in software systems: A quantitative
  perspective.
\newblock In Sushil Jajodia and Duminda Wijesekera, editors, {\em Data and
  Applications Security XIX}, volume 3654 of {\em LNCS}, pages 281--294. 2005.

\bibitem{ANDE-02-OSS} Ross Anderson.
\newblock Security in open versus closed systems - the dance of {Boltzmann,
  Coase and Moore}.
\newblock In {\em Proceedings of Open Source Software: Economics, Law and
  Policy}, 2002.

\bibitem{ARBA-etal-00-IEEE} William~A. Arbaugh, William~L. Fithen, and John
    McHugh.
\newblock Windows of vulnerability: A case study analysis.
\newblock {\em IEEE Computer}, 33(12):52--59, 2000.

\bibitem{AVIZ-etal-04-TDSC} Algirdas Avizienis, Jean-Claude Laprie, Brian
    Randell, and Carl Landwehr.
\newblock Basic concepts and taxonomy of dependable and secure computing.
\newblock {\em IEEE Transactions on Dependable and Secure Computing},
  1(1):11--33, 2004.

\bibitem{DOWD-etal-07} Mark Dowd, John McDonald, and Justin Schuh.
\newblock The art of software security assessment.
\newblock Addision-Wesley publications, 2007.

\bibitem{FLEM-WALL-86-CACM} Philip~J. Fleming and John~J. Wallace.
\newblock How not to lie with statistics: the correct way to summarize
  benchmark results.
\newblock {\em Communication of the ACM}, 29(3):218--221, 1986.

\bibitem{JOH-etal-08-ISSRE} HyunChul Joh, Jinyoo Kim, and Yashwant Malaiya.
\newblock Vulnerability discovery modeling using {W}eibull distribution.
\newblock In {\em Proceedings of the 19th IEEE International Symposium on
  Software Reliability Engineering (ISSRE'08)}, pages 299--300, 2008.

\bibitem{KIM-etal-07-HASE} Jinyoo Kim, Yashwant Malaiya, and Indrajit Ray.
\newblock Vulnerability discovery in multi-version software systems.
\newblock In {\em Proceeding of the 10th IEEE International Symposium on High
  Assurance Systems Engineering}, pages 141--148, 2007.

\bibitem{KRSU-98-PHD} Ivan~Victor Krsul.
\newblock {\em Software Vulnerability Analysis}.
\newblock PhD thesis, Purdue University, 1998.

\bibitem{MASS-etal-11-ESSOS} Fabio Massacci, Stephan Neuhaus, and Viet~Hung
    Nguyen.
\newblock After-life vulnerabilities: A study on firefox evolution, its
  vulnerabilities and fixes.
\newblock In {\em Proceedings of the 2011 Engineering Secure Software and
  Systems Conference (ESSoS'11)}, 2011.

\bibitem{MASS-NGUY-10-METRISEC} Fabio Massacci and Viet~Hung Nguyen.
\newblock Which is the right source for vulnerabilities studies? an empirical
  analysis on mozilla firefox.
\newblock In {\em Proceedings of the International ACM Workshop on Security
  Measurement and Metrics (MetriSec'10)}, 2010.

\bibitem{MCKILLUP-BOOK} Steve McKillup.
\newblock {\em Statistics Explained: An Introductory Guide for Life
  Scientists}.
\newblock Cambridge University Press, 2005.

\bibitem{NGUY-MASS-13-ASIACCS} Viet~Hung Nguyen and Fabio Massacci.
\newblock The (un) reliability of nvd vulnerable versions data: an empirical
  experiment on google chrome vulnerabilities.
\newblock In {\em Proceeding of the 8th ACM Symposium on Information, Computer
  and Communications Security (ASIACCS'13)}, 2013.

\bibitem{NIST-StatBook-12} {NIST/SEMATECH}.
\newblock {\em e-Handbook of Statistical Methods}, 2012.
\newblock http://www.itl.nist.gov/div898/handbook/.

\bibitem{OZMEN-07-QoP} Andy Ozment.
\newblock Improving vulnerability discovery models: Problems with definitions
  and assumptions.
\newblock In {\em Proceedings of the 3rd Workshop on Quality of Protection},
  2007.

\bibitem{RESC-05-SP} Eric Rescorla.
\newblock Is finding security holes a good idea?
\newblock {\em IEEE Security and Privacy}, 3(1):14--19, 2005.

\bibitem{SCHNEIDER-91-NAP} Fred~B. Schneider.
\newblock Trust in cyberspace.
\newblock {\em National Academy Press}, 1991.

\bibitem{SLIW-05-MSR} Jacek Sliwerski, Thomas Zimmermann, and Andreas Zeller.
\newblock When do changes induce fixes?
\newblock In {\em Proceedings of the 2nd International Working Conference on
  Mining Software Repositories MSR('05)}, pages 24--28, May 2005.

\bibitem{WOO-ALHAZMI-MALAIYA-06-SEA} Sung-Whan Woo, Omar Alhazmi, and Yashwant
    Malaiya.
\newblock An analysis of the vulnerability discovery process in web browsers.
\newblock In {\em Proceedings of the 10th IASTED International Conferences
  Software Engineering and Applications}, 2006.

\bibitem{WOO-etal-11-CS} Sung-Whan Woo, HyunChul Joh, Omar Alhazmi, and
    Yashwant Malaiya.
\newblock Modeling vulnerability discovery process in {A}pache and {IIS HTTP}
  servers.
\newblock {\em Computer \& Security}, 30(1):50 -- 62, 2011.

\bibitem{YOUNIS-etal-11-SAM} Awad Younis, HyunChul Joh, and Yashwant Malaiya.
\newblock Modeling learningless vulnerability discovery using a folded
  distribution.
\newblock In {\em Proceeding of the Internaltional Conference Security and
  Management (SAM'11)}, pages 617--623, 2011.

\end{thebibliography}

\appendix
A replication guide of this work could be found online at
\url{https://wiki.science.unitn.it/security/doku.php?id=vulnerability_discovery_models}.
Also, you can find all required materials (\eg tools, scripts, and data) to
rerun the experiment.

\begin{IEEEbiographynophoto}{Viet Hung Nguyen} He is a PhD student in computer science at
University of Trento, Italy under the supervision of professor Fabio Massacci
since November 2009. He received his MSc and BEng in computer science and
computer engineering in 2007 and 2003. Currently, his main interest is the
correlation of vulnerability evolution and software code base evolution.
\end{IEEEbiographynophoto}

\begin{IEEEbiographynophoto}{Fabio Massacci}

\end{IEEEbiographynophoto}

\end{document}